\documentclass[preprint,12pt]{elsarticle}

\newcommand{\etal}{\textit{et al.}}
\newcommand{\revised}[1]{\textcolor{black}{#1}}

\usepackage{amsmath}
\usepackage{tcolorbox}
\usepackage{relsize}
\usepackage{hyperref}
\usepackage{balance}
\usepackage{import}
\usepackage{xcolor,colortbl}
\usepackage{booktabs}
\usepackage{multirow}
\usepackage{paralist}
\usepackage{caption}
\usepackage{xurl}
\usepackage{longtable}
\usepackage[inline]{enumitem}
\DeclareCaptionType{equ}[][]

\begin{document}
\begin{frontmatter}

\title{State-Of-The-Practice in Quality Assurance in Java-Based Open Source Software Development}

\author[1]{Ali Khatami\fnref{fn1}}
\ead{s.khatami@tudelft.nl}
\fntext[fn1]{\href{https://orcid.org/0000-0002-2212-2311}{0000-0002-2212-2311}}

\affiliation[1]{organization={Delft University of Technology},
            addressline={Van Mourik Broekmanweg 6}, 
            city={Delft},
            postcode={2628XE}, 
            country={The Netherlands}}

\author[1]{Andy Zaidman\fnref{fn2}}
\fntext[fn2]{\href{https://orcid.org/0000-0003-2413-3935}{0000-0003-2413-3935}}
\ead{a.e.zaidman@tudelft.nl}

\begin{abstract}
    To ensure the quality of software systems, software engineers can make use of a variety of quality assurance approaches, e.g., software testing, modern code review, automated static analysis, 
    and build automation. Each of these quality assurance practices have been studied in depth in isolation, but there is a clear knowledge gap when it comes to our understanding of how these 
    approaches are being used in conjunction, or not. In our study, we broadly investigate whether and how these quality assurance approaches are being used in conjunction in the development 
    of 1,454 popular open source software projects on GitHub. Our study indicates that typically projects do not follow all quality assurance practices together with high intensity. In fact, we only 
    observe weak correlation among some quality assurance practices. In general, our study provides a deeper understanding of how existing quality assurance approaches are currently being 
    used in \revised{Java-based} open source software development. 
    \revised{Besides, we specifically zoomed in on the more mature projects in our dataset, and generally we observe that more mature projects 
    are more intense in their application of the quality assurance practices, 
    with more focus on their ASAT usage, and code reviewing, but no strong change in their CI usage.}
\end{abstract}

\begin{keyword}
    Software Quality Assurance, Software Testing, Continuous Integration, Code Review, Build Automation, Automated Static Analysis
\end{keyword}

\end{frontmatter}

\section{Introduction}\label{section:introduction}
We have grown accustomed to living in a software-filled world: we rely on our smartphone full of apps, or on digitalised services for many of our daily activities~\cite{TechCrunch2016}. 
With software pervading modern society and controlling critical aspects from an economic, safety, security and scientific standpoint, the quality and dependability of that software is indispensable~\cite{JazayeriASE2004}, 
because of the potential catastrophic consequences~\cite{DBLP:conf/icse/KoDD14}.
To ensure the quality of these software systems, software engineers can use a range of quality assurance approaches, e.g., software testing~\cite{Aniche2022,DBLP:journals/tse/KameiSAHMSU13}, modern code review~\cite{DBLP:conf/icse/BacchelliB13,DBLP:journals/tse/KameiSAHMSU13}, automated static analysis~\cite{DBLP:conf/wcre/BellerBMZ16,DBLP:journals/ese/VassalloPPPGZ20}, and build automation~\cite{DBLP:conf/msr/BellerGZ17,RahmanRCOSE2017}.

Each of these quality assurance techniques have been studied in isolation in-depth. For example, Beller \etal~have analyzed the prevalence of automated static analysis tools in open source software development~\cite{DBLP:conf/wcre/BellerBMZ16}, Rausch \etal~performed an empirical analysis of build failures~\cite{DBLP:conf/msr/RauschHL017}, both Beller \etal~and Rigby \etal~investigated aspects of code reviews in open-source software development~\cite{DBLP:conf/msr/BellerBZJ14,DBLP:conf/icse/RigbyGS08}, and Hilton \etal~focused on understanding continuous integration~\cite{DBLP:conf/kbse/HiltonTHMD16}. 
Other studies have examined the relationship between two quality assurance practices, e.g., Cassee \etal~have studied the impact of continuous integration on code review~\cite{DBLP:conf/wcre/CasseeVS20}, Zhao \etal~have looked into the impact of continuous integration on other software development practices~\cite{DBLP:conf/kbse/ZhaoSZFV17}, Zampetti \etal~did a study on the interplay between pull requests and CI~\cite{DBLP:conf/wcre/ZampettiBCP19}, Panichella \etal~looked into the complementarity of static analysis tools and code reviews~\cite{DBLP:conf/wcre/PanichellaAPA15}, and Nery \etal~did an empirical study of the relationship between CI and test code evolution~\cite{DBLP:conf/icsm/NeryCK19}.

Some of these approaches are to be considered complimentary to each other, e.g., code reviews and software tests find different issues~\cite{MantylaTSE2009}. Precisely because of the complementarity between the approaches, we set out to broadly investigate whether and how software testing, modern code reviews, automated static analysis, and build automation are being used in conjunction in the development of open source software on GitHub.
Github is one of the most important development platforms for open source software and has around 28 million users and 79 million repositories~\cite{BorgesJSS2018}. In addition to it being a git-based version control system, GitHub integrates several elements for quality assurance, e.g., pull requests for code review, continuous integration for build automation and test execution, and integrations for automated static analysis and code coverage. 

We have performed our investigation into the state-of-the-practice in quality assurance by analysing \textbf{1454} GitHub projects. Our investigation is guided by the following research questions: 
\begin{itemize}
\item[\textbf{RQ1}] What is the current state-of-the-practice in quality assurance in open source software development? 

\revised{The main goal of our study is to see how quality assurance practices are being followed in conjunction. Before we can study how they are used in 
conjunction, we need to understand how they are being used in isolation. We fully acknowledge that each of the quality assurance practices has previously 
been studied in isolation, however, we find it essential to understand the \emph{current} state-of-practice reflected in our dataset that comes from GitHub.
We see the study of quality assurance practices in isolation as a cornerstone to build towards studying them in conjunction; we do not claim the study in isolation as a
contribution of our work.
} 
	\begin{itemize}
		\item[\textbf{RQ1.1}] What is the prevalence of quality assurance approaches like software testing, modern code review, automated static analysis, and buildability?

		\item[\textbf{RQ1.2}] Which quality assurance approaches are being used in conjunction?
		
	\end{itemize}
		\revised{Our set of selected projects consist of popular and active projects on GitHub (1454 in total). However, we hypothesize that projects that have been more active 
		and are relatively more popular than others follow quality assurance practices with higher intensity. Thus, we aim to see how prevalence of quality assurance 
		practices differs in the top 50 most popular and active projects compared to all the projects in our dataset.} 
	\begin{itemize}	
		\item[\textbf{RQ1.3}] \revised{How do mature projects follow quality assurance practices?}
	\end{itemize}
\end{itemize}

\smallskip
\revised{While conducting our research, we have faced a number of challenges studying each of the practices, e.g., it was not easy to build all projects out of the box. 
Further on in this paper, we provide an overview of these challenges and hypothesize how they are important not only for researchers, but also for practitioners.} 

\smallskip
Getting a broad understanding of how existing quality assurance approaches are being used in open source development is important for software practitioners, software engineering educators, and software engineering researchers. 
Specifically, this understanding will inform practitioners about the (combinations of) best practices and how they are currently in use. Furthermore, it will apprise educators on what should be taught in software engineering classes, 
both in terms of the current state-of-practice and how we can improve on the state-of-practice. Finally, it can inspire researchers to improve both quality assurance approaches in isolation, as well as how their combinations can be strengthened.

The structure of the remainder of this paper is as follows. Section~\ref{section:background} presents the related literature. Section~\ref{section:studysetup} details our data collection procedure. Section~\ref{section:results} reports on our results for each of the quality assurance practices in isolation and in conjunction. 
Section~\ref{section:discussion} discusses our findings, while Section~\ref{section:conclusion} concludes the paper.

There is also an associated package provided to replicate and reproduce the results of our study~\cite{ali_khatami_2022_7404903}.

\section{Background} \label{section:background}
We present literature related to each of the practices in our study as the background knowledge in this section.

\subsection{‌(CI) Building, Compilability}

Tufano \etal~have examined the compilability of $\sim$220,000 snapshots of 100 
Apache Software Foundation projects~\cite{DBLP:journals/smr/TufanoPBPOLP17}. 
They found that 30\% (median) of those snapshots could not be compiled. 
\revised{Maes-Bermejo \etal~revisited Tufano \etal's work by replicating their study and 
also reproducing a different set of 80 projects with $\sim$300,000 snapshots~\cite{DBLP:journals/ese/Maes-BermejoGGR22}. 
They found that the most influential error causing build failures are missing external artifacts, 
contradicting Tufano \etal's finding that dependency-issues is the most common cause of build failures.}
Hassan \etal~have investigated Java projects on GitHub and report that 101 out of 187 projects could be successfully build using Ant, Maven, or Gradle~\cite{DBLP:conf/esem/HassanMLW17}. Their manual inspection provides indications on 
the main reasons for build failure: `backward-incompatibility of JDK and building tools`, `non-default parameters in build commands`, and `project defects in code / configuration files`. 

In another study Hassan \etal~have proposed a model to predict CI build results~\cite{DBLP:conf/esem/HassanW17}. More specifically, they have built a classifier using the random forest algorithm and applied it to the TravisTorrent dataset~\cite{DBLP:conf/msr/BellerGZ17a} of 402 Java projects and 250,000 builds. 

Jin and Servant have proposed an approached called SmartBuildSkip to reduce CI build cost by predicting successful CI builds and skipping them~\cite{DBLP:conf/icse/JinS20}. They base their approach on two hypotheses: most builds pass, and build failures usually happen consecutively (52\% of build failures are consecutive). 

Lou \etal~have studied 1,080 questions on Stack Overflow related to build issues of Maven, Gradle and Ant build systems~\cite{DBLP:conf/sigsoft/LouCCHZ20}.  
They have observed a high diversity in build failure categories, however about 68\% of them can be fixed by applying fix patterns in the build script code related to plugins and dependencies (the most frequent ones being (1) correcting the plugin setting, and (2) adding missing dependencies).  

Rausch \etal~have analyzed build failures in CI from 14 open-source Java projects linking commits to their CI builds results~\cite{DBLP:conf/msr/RauschHL017}. As a result, they found 14 common error categories and reported test failures as the most common one among projects. 

Zampetti \etal~empirically investigated how developers use CI builds outcome during a code review in a PR discussion, analyzing 69 projects using Travis-CI on GitHub~\cite{DBLP:conf/wcre/ZampettiBCP19}. They found that passed builds increase the chances of merging a PR, and if a build fails, process-factors (like the age of PR, changed lines and files, review comments, etc.) have stronger correlation. 
Interestingly, sometimes a PR is merged despite a failing build status, e.g., due to minor warnings from static analysis tools.

\subsection{Code Review}
\revised{McIntosh \etal~have provided empirical evidence that higher code review participation and lower coverage in reviews
will lead to less defect-prone, higher quality software. In their study they considered post-release defects as a proxy for 
software quality. Since their study was conducted on Gerrit\footnote{\url{https://www.gerritcodereview.com/}, last visited Nov 23rd, 2022.} (a web-based code review tool for git-based software projects) and not on GitHub, we could not use their same metrics to measure 
code review participation in our projects. Moreover, their study is focused on the impact of code review coverage and code review participation 
on software quality, whereas we study the relation between code review intensity and other quality assurance practices~\cite{DBLP:conf/msr/McIntoshKAH14}.}

Cassee \etal~have explored code reviews of 685 open source projects on GitHub that used Travis-CI as their CI platform~\cite{DBLP:conf/wcre/CasseeVS20}. They found that using Continuous Integration (CI) on average saves one review comment per pull request, indicating that 
adopting continuous integration can save-up valuable time in code review. 

Similarly, Rahman and Roy have investigated the impact of CI on code reviews to understand the relation between CI builds and the participation or quality of the code reviews~\cite{DBLP:conf/msr/RahmanR17}. They indicate that succeeding automated builds have a positive effect on code review participation. Using code review comments count as a proxy for code review quality, they came to the conclusion that automated build frequency has a significant positive impact on the code review quality.

Rigby \etal~ have conducted a case study of the Apache Server peer review practices as a popular open source project~\cite{DBLP:conf/icse/RigbyGS08}. They observe the number of reviewers that are involved in each review, the frequency of reviews, the size of the artifacts under review, the time interval of reviews, and how many reviews find defects. 

Beller \etal~have investigated to see what problems are fixed when doing modern code reviews in open source projects~\cite{DBLP:conf/msr/BellerBZJ14}.  
They report that 10 to 35\% of review suggestions do not lead to any code changes, and they observe a 3-to-1 ratio in maintainability versus functional code review comments. 

Bacchelli and Bird coined the term \textit{Modern Code Review (MCR)}~\cite{DBLP:conf/icse/BacchelliB13} as the lightweight and less formal version of  heavyweight formal code inspections~\cite{DBLP:journals/ibmsj/Fagen76}. They explored the expectations, outcomes, and challenges of MCR by manually classifying comments, and observing, interviewing, and surveying practitioners. Their study found that the main motivation of code reviews is not to find defects/bugs, but to increase awareness and do knowledge transfer. 

Kononenko \etal~have looked into developers' perception of code review quality~\cite{DBLP:conf/icse/KononenkoBG16}. They found that review quality is primarily associated with thoroughness of the feedback, familiarity with the code, and the quality of the code.  
Factors such as code quality, presence and quality of tests, and personality of the developer would have impact on the review decision. 

\subsection{Testing}
\revised{The purpose of software testing is to improve the quality of software by the detection of 
faults~\cite{DBLP:conf/icse/GopinathJG14}, which makes testing a good candidate among the 
quality assurance practices that we consider in our study. 
}
Hilton \etal~have performed a large scale study of test coverage evolution over 7,816 builds of 47 open source projects~\cite{DBLP:conf/kbse/Hilton0M18}. 
To gather coverage data of projects they ran test suits themselves and also used the Coveralls service\footnote{\url{https://coveralls.io}, last visited May 20th, 2022.} for some of them. 
Their main findings are that patch coverage is not correlated with overall coverage, and that patches might have a significant effect on overall coverage. 
An observation that was previously also made by Elbaum \etal~\cite{DBLP:conf/icsm/ElbaumGR01}.

Zaidman \etal~have observed that testing activities do not always nicely co-evolve with production code engineering 
activities~\cite{DBLP:journals/ese/ZaidmanRDD11}. Instead, they observe that testing happens in bursts. 
Importantly, this observation was made before the pull-based development model became popular. 
Gousios \etal~have deeply studied the pull-based development contribution model and found that guidelines 
for pull request creation include feature isolation, but also conformance to source code quality and test coverage 
guidelines~\cite{DBLP:conf/icse/GousiosZSD15}.

Beller \etal~have tracked how developers go about the test engineering process in their Integrated Development Environment 
(IDE)~\cite{DBLP:journals/tse/BellerGPPAZ19}. From the software engineers that they tracked, the found that around 
50\% of them did not carry out any test-related activities (writing or executing test) in the IDE during a 5-month period.  

\revised{
Kochhar \etal~have found in their study 
that the relationship between test coverage and post-release 
bugs is either non-existent or unclear \cite{DBLP:journals/tr/KochharLLN17}. However, 
in another study by Athanasiou \etal,~on the relation between test code quality and 
issue handling performance, they have proposed a test code quality model, and have 
used code coverage as a metric to assess test code's completeness. They have also showed 
that there is a positive correlation between test code quality and throughput and productivity of 
issue handling~\cite{DBLP:journals/tse/AthanasiouNVZ14}. 
}

\subsection{Continuous Integration}

Vasilescu \etal~have investigated the effects of adopting continuous integration on quality of the software and also on the productivity of the development team~\cite{DBLP:conf/sigsoft/VasilescuYWDF15}. 
They report that using CI leads to more pull requests getting accepted, and that CI 
improves the productivity of project teams.

Vassallo \etal~have investigated the concept of \textit{Continuous Code Quality} (CCQ) that they take from SonarQube~\cite{SonarSource}. CCQ is the idea that CI stands central in ensure code quality by running automated tests and performing automated code inspections at every build~\cite{DBLP:conf/kbse/VassalloPBG18}. Their main finding after a large-scale analysis of 119 Java projects is that only 11\% of CI builds continuously monitor code quality. 

Hilton \etal~have studied the usage, costs, and benefits of CI in open-source projects~\cite{DBLP:conf/kbse/HiltonTHMD16}. 
They observe that popular projects are more likely to use CI, projects using CI release more often than the ones not using it, projects use CI because it helps them catch bugs early. 

Nery \etal~have looked into the relationship between continuous integration and test code evolution~\cite{DBLP:conf/icsm/NeryCK19}. They study 
test coverage and test ratio (the proportion of test code in a project) and report that projects using CI have a rising trend in test ratio compared to projects that do not use CI, and that test coverage of CI-using projects grows more overtime compared to projects not using CI. 

\revised{Hilton \etal~have studied the barriers of CI usage, the needs of developer using CI tools, 
their motives, and the benefits they experienced when using CI tools~\cite{DBLP:conf/sigsoft/Hilton0TMD17}. In their study, developers 
mentioned the need of user interfaces for modifying CI configurations, and easier configuration of CI services.
Developers also mentioned maintaining and setting up CI services as barriers of CI usage. It is worth mentioning that 
this study was conducted before the introduction of GitHub Actions as a CI platform on GitHub. Since GitHub Actions claims 
to make the automation of software workflows easier, we are interested to see the prevalence of today's continuous integration usage
by studying Java projects on GitHub more than 2 years after the introduction of GitHub Actions.}

\revised{Gautam \etal~have used the TravisTorrent dataset~\cite{DBLP:conf/msr/BellerGZ17a} to conduct an empirical study 
on open source projects following continuous integration. They focused on reaching a good understanding of the practices 
followed by software projects to attract talents and help on-boarding new members~\cite{DBLP:conf/msr/GautamVS17}.
}

\subsection{Other Elements of Quality Assurance}

Gousios \etal~defined pull-based software development model~\cite{DBLP:conf/icse/GousiosPD14}. In their exploratory analysis on Github projects at that time they discovered that pull-based model is not popular. Moreover, they found that most pull requests are small (affected a few dozen lines of code), and only a few (13\%) of them are rejected due to technical reasons. Furthermore, they reported that providing guidelines for pull request processing, and having a high coverage test suite will attract external contributors and speed up merging of contributions.

Gousios \etal's study~\cite{DBLP:conf/icse/GousiosPD14} resulted in creation of a dataset of 900 projects and 350,000 pull requests called pullreqs~\cite{DBLP:conf/msr/GousiosZ14}. This dataset was initially created by Gousios and Zaidman, then updated by Zhang \etal ~\cite{DBLP:conf/msr/ZhangR020} by adding more projects, pull requests, and data features. In both studies there are similar data features that overlap with features collected in our study, using similar heuristics. Besides, heuristics used in Zhang \etal~work for collecting CI usage, and CI build data are similar to our data collection methodology.

Kinsman \etal~have examined how software developers use GitHub Actions~\cite{DBLP:conf/msr/KinsmanWGT21}. GitHub Actions is a GitHub feature to automate tasks based on events like commits, PRs, issues, etc. In their dataset only 0.7\% of projects adopted GitHub Action spread across 20 categories (CI, CD, utilities, etc.).

\section{Study Setup} \label{section:studysetup}
\begin{figure*}
    \centering
    \includegraphics[width=0.95\textwidth]{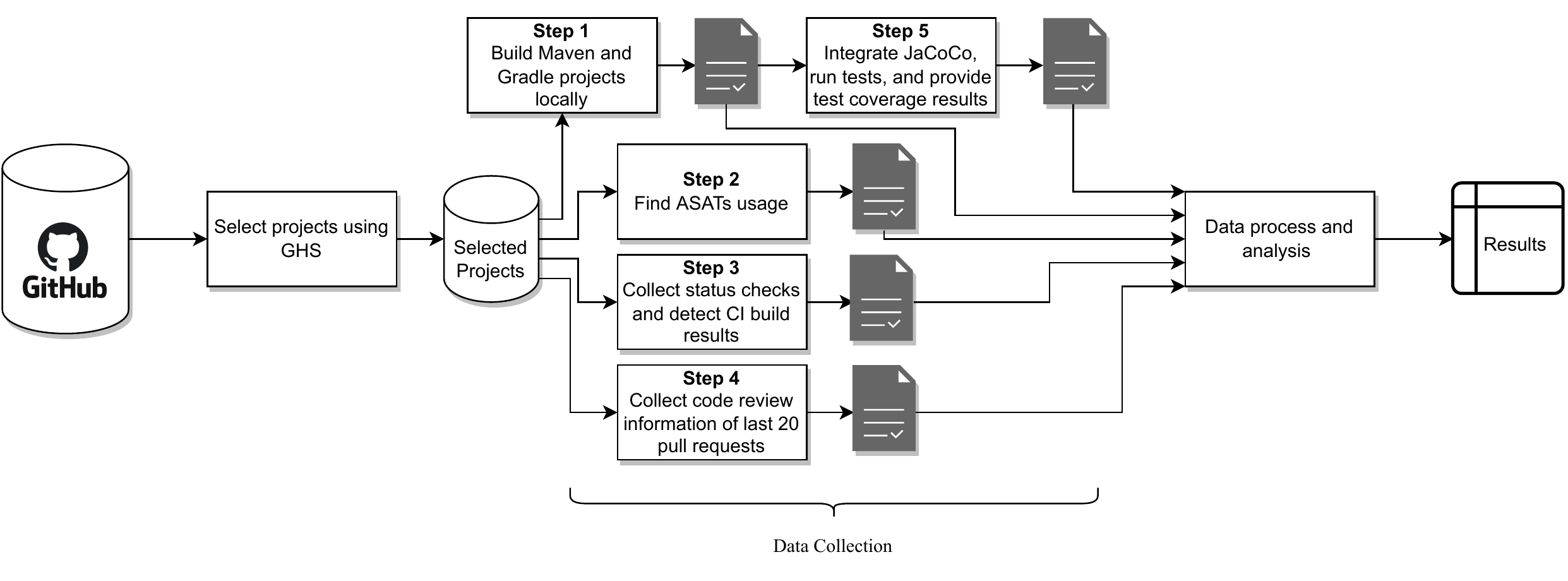}
    \caption{An overview of data collection steps.}
    \label{fig:data-collection-overview}
\end{figure*}

In this section, we first explain how we selected projects for the study, 
and then describe the data collection steps in detail. \revised{Futhermore, we show how 
we selected the mature projects out of the main dataset.}
Figure~\ref{fig:data-collection-overview} details our data collection 
procedure in 5 steps, each for a specific quality assurance practice. 
\begin{compactdesc}
\item[Step 1:] Build projects locally.
\item[Step 2:] Determine usage of Automated Static Analysis Tools (ASATs).
\item[Step 3:] Investigate CI usage and build result.
\item[Step 4:] Collect code review data from last 20 pull requests.
\item[Step 5:] Collect test coverage results.
\end{compactdesc}

\subsection{Studied Projects}

We used Dabic \etal's GitHub Search (GHS) dataset~\cite{Dabic:msr2021data} to select projects from GitHub based on different criteria. As it is our aim to investigate projects that are likely to be in a position to make optimal use of quality assurance practices, we filtered the projects using the following criteria. We selected popular ($>$100 stars) Java projects that have recently been active (last commit after August 2021). We have settled on one programming language to keep our data analysis pipeline simple, more specifically, we have chosen Java. We have also selected projects to minimally have 10 contributors, as our line of reasoning is that you need to have a sufficient number of developers to make optimal use of quality assurance practices like code reviews. In addition, we haven chosen to only study non-forked projects to avoid duplicate projects. Finally, we put a lower limit of 200 pull requests for a project to be part of our data set; this threshold  ensures that we study projects that use the pull based development model, and are thus potentially also doing code reviews on the GitHub platform. 

We also saved the latest commit hash for each project, and in the next steps, we use this commit hash to run git checkout. This enables (1) us to study each of the projects in a specific state, and (2) to easily replicate our work in future~\cite{ali_khatami_2022_7404903}.

\subsection{Data Collection}
In this section we go step by step and explain the path we took to collect data related to each quality assurance practice.

\subsubsection{\textbf{Step 1. Local builds}} \label{local-builds-data}
In this step we describe how we have collected information on the buildability (compilability) of projects. This information first shows us which projects can be built out of the box (without any special configuration, reading project specific documents, and environment setup), only using a minimal environment (a Docker container of Ubuntu 21.04) with the Java Development Kit (Open JDK 11) and build tools such as Maven and Gradle. Secondly, we will use this information to filter successfully built projects and generate test coverage results for them in Step 4.

Starting from the collection of GitHub Java projects, first we filtered the projects using Maven and Gradle as their build system. To do this we searched for \textit{gradle.build} and \textit{pom.xml} configuration files inside (the root path of project) repositories by querying GitHub. 
Next, we specifically filtered out 
Android projects (by searching for the \textit{AndroidManifest.xml} file inside repositories) to avoid build failures related to Android SDK configurations.

After removing non-Gradle, non-Maven, and Android projects, we obtain the set of projects to be built locally. In the next step, we automatically clone projects and after doing git checkout on the latest commit hash, build projects locally using a docker container with \textit{Ubuntu 21.04}, \textit{OpenJDK v11}, \textit{Gradle v7.1}, \textit{Python v3.9}, and \textit{Maven v3.8.1} installed. Similar to Sul{\'{\i}}r \etal~\cite{DBLP:journals/corr/abs-1712-01024} we skip running tests when building projects in this step to avoid test failures. We add relevant flags to our build commands to have full log information in build output logs (see Table~\ref{tab:1}). Furthermore, we put one hour maximum build execution time to avoid infinite builds (similar to Sul{\'{\i}}r \etal~\cite{DBLP:journals/corr/abs-1712-01024}).

\begin{table}[!t]
\centering
\begin{smaller}
\begin{tabular}{ll}
    \toprule
    \textbf{Build System} & \multicolumn{1}{c}{\textbf{Build Command}}    \\ \midrule
    Gradle                & \texttt{./gradlew clean assemble --stacktrace}         \\
    Maven                 & \texttt{\small mvn clean package -DskipTests --batch-mode -e} \\ \bottomrule
\end{tabular}
\end{smaller}
\caption{\label{tab:1}Build commands for different build tools.}
\end{table}

\subsubsection{\textbf{Step 2. Automatic static analysis tools}} \label{sat-data}
Similar to the previous step, we again use GitHub's API to see if there is a configuration file related to one of the more popular Java static analysis tools (SpotBugs, Checkstyle, and FindBugs) in the repositories, similar to Beller~\etal~\cite{DBLP:conf/wcre/BellerBMZ16}. To do this, after running the git checkout on the latest commit hash, we looped through the repository Git tree to check whether it contains a file with a substring of \textit{`spotbugs'} in their name phrase. If that is the case, then the project is marked as (likely) to use the SpotBugs tool. We performed a similar step for Checkstyle and FindBugs. 

\subsubsection{\textbf{Step 3. Continuous integration}} \label{ci-data}
In this step, we collected the required data to study both the prevalence of CI and the results of the build steps that are part of CI. 
For the prevalence of CI, we need CI usage data (whether a project is doing continuous integration). In terms of the result of CI, we are interested in two distinct steps: (1) the overall CI build success, and (2) the result of specifically the build phase of the CI workflow. We need the latter result, to be able to compare the result of the CI build phase, to that of a local build result from Section~\ref{local-builds-data}.

\textbf{Step 3.1. CI usage}: 
To establish whether a project uses continuous integration, we have decided to use a different approach to earlier studies on this topic. In particular, prior studies have used the TravisTorrent dataset~\cite{DBLP:conf/msr/BellerGZ17a}, or the Travis-CI API to check whether a project is using CI~\cite{DBLP:conf/wcre/CasseeVS20}. Also, Vassallo \etal's work only considered SonarCloud as a CI platform to study projects' code quality~\cite{DBLP:conf/kbse/VassalloPBG18}, and Gallaba \etal's study was limited to CircleCI builds~\cite{gallaba2022lessons}. 
These aforementioned approaches are limited to one particular CI platform, namely Travis-CI, SonarCloud, or CircleCI. To consider different CI platforms together, we followed a different route to establish CI usages, namely the usage of \emph{checks} on GitHub\footnote{\url{https://docs.github.com/en/pull-requests/collaborating-with-pull-requests/collaborating-on-repositories-with-code-quality-features/about-status-checks}, last visited May 20th, 2022.}.

\begin{figure*}
    \centering
    \includegraphics[width=0.95\textwidth]{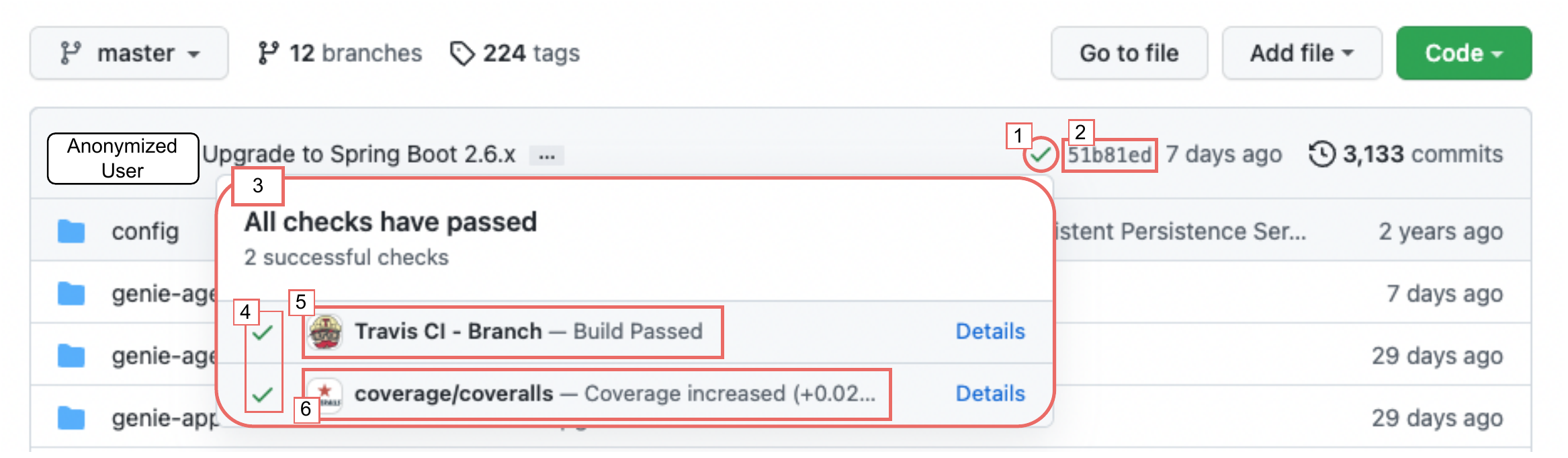}
    \caption{Status checks on a commit in the main page of a repository: 1. The combined result of all status checks (in this case successful), 2. Related commit hash, 3. List of all status checks and a summary of their results, 4. Each status check's result, 5. A check by a GitHub App (TravisCI), 6. A status checked marked by an external service (Coveralls), these status checks are called ``statuses''}
    \label{fig:checks-github}
\end{figure*}

Using GitHub Actions\footnote{\url{https://docs.github.com/en/actions}, last visited May 20th, 2022.} (GHA) projects can automate their workflows, including CI/CDs on GitHub. Events like pushing a commit, creating a pull request, opening an issue, etc. can trigger a workflow run (also referred to as a \textit{check}). In this paper we collect information related to \textit{checks} that run after pushing a specific commit, namely latest commit SHA of projects in our dataset.

There are two types of \textit{checks} that can run after pushing a commit\footnote{\url{https://docs.github.com/en/pull-requests/collaborating-with-pull-requests/collaborating-on-repositories-with-code-quality-features/about-status-checks}, last visited May 20th, 2022.}: (1) the ones defined by workflows, which are only available for use with GitHub Apps, and (2) the ones that are marked with external services through the GitHub API (these are called \textit{Commit Statuses} in GitHub's documentation), shown in Figure~\ref{fig:checks-github}. From now on, similar to GitHub's documentation, we will call the first type: ``checks'', the second type: ``(commit) statuses'', and both together: ``status checks''.

What is important to note is that not all status checks are related to CI, and as such can not be unequivocally regarded as a sign of CI usage. To address this, we heuristically filtered CI related status checks. Table~\ref{tab:checks-attributes} shows data attributes of status checks depending on their type. This information is collected from the GitHub GraphQL API\footnote{\url{https://docs.github.com/en/graphql}, last visited May 20th, 2022.}.

To filter CI related status checks, depending on their type, we look for some keywords in \textit{context}, \textit{targetUrl}, and \textit{description} of statuses, and for checks we look for usage of specific GitHub Apps related to CI.  
For GitHub Actions App\footnote{GitHub Actions App is a GitHub App and differs from GitHub Actions which is GitHub's CI/CD platform. In this paper, with GHA we refer to the CI/CD platform on GitHub, and with GHAA we refer to the GitHub Actions App on GitHub's marketplace.} (GHAA) checks, the most used GitHub App, we have also looked to see if they contain any ``build'' keyword in their \textit{name}. This is because using GHAA does not always mean a project is using CI, however if there is the build keyword, there is a strong signal of the project doing CI. Table~\ref{tab:checks-filter-keywords} shows the keywords used in filtering CI related status checks.

\begin{table}[!t]
    \centering
    \renewcommand{\tabcolsep}{0.15cm}
\begin{smaller}
\begin{tabular}{lp{5cm}p{5.8cm}}
    \toprule
    \multicolumn{3}{c}{\textbf{Checks' Attributes}}                                                                                                                                                                                                                                                         \\
    \textbf{Name} & \textbf{Description}                                                                                                   & \textbf{Value(s)}                                                                                                                                              \\ \hline
    conclusion    & The conclusion of the check run.                                                                                       & ACTION\_REQUIRED / CANCELLED /  FAILURE / NEUTRAL / SKIPPED /  STALE / STARTUO\_ FAILURE /  SUCCESS / TIMED\_OUT\\ \hline
    name          & The name of the check for this check run.                                                                              & String                                                                                                                                                         \\ \hline
    status        & The current status of the check run.                                                                                   & COMPLETED / IN\_PROGRESS / PENDING / QUEUED / REQUESTED / WAITING                                                \\ \hline
    summary       & Summary of the check run.                                                                                              & String                                                                                                                                                         \\ \hline
    text          & Text of check run.                                                                                                     & String                                                                                                                                                         \\ \hline
    title         & Title of check run.                                                                                                    & String                                                                                                                                                         \\ \hline
    app           & The GitHub App which created the check suite containing this check run.     & \begin{tabular}[c]{@{}l@{}}Name: String\\ Slug: String          \end{tabular}                                                                                   \\ \hline
    workflow run  & The workflow run associated with the check suite containing this check run. & \begin{tabular}[c]{@{}l@{}}Number: Int\\ URL: URI\end{tabular}                                                                                                 \\ \hline
    workflow      & The workflow executed in the workflow run.                                                                             & Name: String                                                                                                                                                   \\ \bottomrule
    \\
    \\
    \toprule
    \multicolumn{3}{c}{\textbf{Statuses' Attributes}}                                                                                                                                                                                                                                                       \\
    \textbf{Name} & \textbf{Description}                                                                                                   & \textbf{Value(s)}                                                                                                                                              \\ \hline
    context       & The name of this status.                                                                                               & String                                                                                                                                                         \\ \hline
    description   & The description for this status.                                                                                       & String                                                                                                                                                         \\ \hline
    state         & The state of this status.                                                                                              & \begin{tabular}[c]{@{}l@{}}ERROR / EXPECTED / FAILURE /\\ PENDING / SUCCESS\end{tabular}                                                                       \\ \hline
    targetUrl     & The URL for this status context.                                                                                       & URI                                                                                                                                     \\ \bottomrule                      
\end{tabular}
\end{smaller}
\caption{\label{tab:checks-attributes}Data attributes collected for status checks from GitHub GraphQL API.}
\end{table}

\begin{table}[!t]
    \centering
\begin{smaller}
    \begin{tabular}{l|l|l}
        \toprule
        \textbf{Type}           & \textbf{Attribute} & \textbf{Keywords}                                                                                                                                                       \\ \hline
        Check                   & app name           & \begin{tabular}[c]{@{}l@{}}GitHub Actions, Azure Pipelines, ci.jenkins.io, \\ ASF Cloudbees Jenkins ci-builds, Travis CI, \\ CircleCI Checks, Cirrus CI, Confluent-Jenkins-CI\end{tabular} \\ \hline
        \multirow{3}{*}{Status} & context            & ci, continuous-integration, kokoro, bazel                                                                                                                                                  \\ \cline{2-3} 
                                & description        & teamcity, kokoro                                                                                                                                                                           \\ \cline{2-3} 
                                & targetUrl          & ci., screwdriver, buildkite, teamcity 
\\ \bottomrule
\end{tabular}
\end{smaller}
\caption{\label{tab:checks-filter-keywords}Keywords used to filter CI related status checks based on their type.}
\end{table}

\textbf{Step 3.2. Overall CI results}: 
To see what is the final state of CI workflows run in a specific state of the project (latest commit hash), we rely on the status checks information. 
Status checks data collected from the GitHub API contains a status check attribute that represents the rollup for both checks and statuses for a commit\footnote{\url{https://docs.github.com/en/graphql/reference/objects\#statuscheckrollup}, last visited May 20th, 2022.} (also referred to as combined status\footnote{\url{https://docs.github.com/en/rest/commits/statuses\#get-the-combined-status-for-a-specific-reference}, last visited May 20th, 2022.}, however only for statuses). We also refer to this rollup result as the CI result of each project for the latest commit hash in our dataset. Of course this only applies to the projects having at least one filtered CI related status check.

\textbf{Step 3.3. Result of build phase in the complete CI workflow}: 
Having local build results in hand from the first step of our data collection (see Section~\ref{local-builds-data}), we want to see how a CI build outcome of a project matches its local build outcome. Therefore, we first heuristically filter build related CI status checks by searching for ``build''/``built'' keywords in their attributes (Table~\ref{tab:checks-build-filter-keywords} shows the details). 
Second, based on the results of each of filtered CI builds, we determine the rollup result of CI builds. Meaning if all statuses' \textit{state} and all checks' \textit{conclusion} is ``SUCCESS'', then the rollup status of CI build is success, otherwise it is failure. 

\begin{table}[!t]
\centering
\begin{smaller}
\begin{tabular}{l|l|l}
\toprule
    \textbf{Type}           & \textbf{Attribute} & \textbf{Keyword(s)} \\ \hline
    \multirow{3}{*}{Check}  & title              & built, build        \\ \cline{2-3} 
                            & text               & build               \\ \cline{2-3} 
                            & name               & build               \\ \hline
    \multirow{2}{*}{Status} & context            & build               \\ \cline{2-3} 
                            & description        & built, build       
\\ \bottomrule
\end{tabular}
\end{smaller}
\caption{\label{tab:checks-build-filter-keywords}Keywords used to filter CI builds status checks based on their type.}
\end{table}

\subsubsection{\textbf{Step 4. Code review}} \label{code-review-data}
To collect information about projects' code review activity, we focus on the user's code review behavior during pull-based development in the GitHub environment. When a contributor creates a pull request to merge their contribution to the main repository, integrators or co-contributors can review the pull request~\cite{DBLP:conf/icse/GousiosZSD15}. 
The contributor who created the pull request and the code reviewer(s) can have a conversation through comments. For our dataset we count the number of reviews and the number of comments for the last 20 merged pull requests of projects (reference data August 1st, 2021). 
We set the specific reference date to ensure future reproducibility. Reviews are defined as pull request review comments with an optional body comment and a state\footnote{\url{https://docs.github.com/en/rest/reference/pulls\#reviews}, last visited May 20th, 2022.}. Approving changes are also considered as a review action. 
With comments, we mean other comments in a pull request discussion part that are not posted by GitHub bots~\cite{wesselEMSE}, and are not made on a unified diff and are usually meant for general communication. We analyse the combination of these two data points to show how code review is taking place for a project on GitHub.

We only consider merged pull requests (based on the GitHub API\footnote{\url{https://docs.github.com/en/rest/reference/pulls\#check-if-a-pull-request-has-been-merged}, last visited May 20th, 2022.}). 
To avoid getting into a loop of checking all pull requests to see if they are merged or not, 
we limit ourselves to the last 100 pull requests. 
Important to note is that if a project is using Git instead of GitHub to merge pull requests~\cite{DBLP:conf/msr/KalliamvakouGBSGD14}, it will not be considered for code reviewing in our dataset, since our approach only monitors code reviews performed through the GitHub web interface.

Furthermore, for our correlation analysis in Section~\ref{section:put-together} we collected two other data points for each of the pull requests: total count of lines changed and count of files changed in all commits. We have combined them to formulate our code review rate metric (Equation~\ref{equ:code-review-rate}). 
We settled on this equation for two key reasons. The first one being that we want to include both count of comments and count of reviews, as they are both important during the review process. The second reason being that we want to normalise the combined count of comments and reviews over the number of changed lines of code divided by the number of changed files. This normalisation enables us to more easily compare the review intensity of a changed piece of source code.

\begin{equ}[!ht]
    \begin{equation}
        \frac{\#Reviews + \#Comments}{\frac{\#TotalChangedLines}{\#TotalChangedFiles}}
        \label{equ:code-review-rate}
    \end{equation}
    \caption*{Equation 1: Code review intensity metric formula.}
\end{equ}

\subsubsection*{\revised{\textbf{Validation of the code review intensity metric}}}
\revised{We carried out a validation of the code review intensity metric by investigating pull requests 
that show a wide variety in code review intensity scores. In particular, we look at code review intensity scores that lie at the 20th, 40th, 60th, and 80th percentile; 
we also consider the minimum and maximum code review intensity scores. We then investigate pull requests 
of 6 projects for a deeper comparison, in which we also checked the discussions happening during code review to better gauge the intensity of
the code reviewing happening.}

\begin{longtable}{l|l|l|l|l|l|l|l|l|l}
\toprule
\multicolumn{1}{c|}{\textbf{\begin{tabular}[c]{@{}c@{}}\rotatebox{90}{Project Name}\end{tabular}}} & 
\multicolumn{1}{c|}{\textbf{\begin{tabular}[c]{@{}c@{}}\rotatebox{90}{Code Review Intensity}\end{tabular}}} & 
\multicolumn{1}{c|}{\textbf{\begin{tabular}[c]{@{}c@{}}\rotatebox{90}{Percentile}\end{tabular}}} & 
\multicolumn{1}{c|}{\textbf{\begin{tabular}[c]{@{}c@{}}\rotatebox{90}{PR-ID}\end{tabular}}} & 
\multicolumn{1}{c|}{\textbf{\begin{tabular}[c]{@{}c@{}}\rotatebox{90}{Change Requested}\end{tabular}}} & 
\multicolumn{1}{c|}{\textbf{\begin{tabular}[c]{@{}c@{}}\rotatebox{90}{\# Reviewers}\end{tabular}}} & 
\multicolumn{1}{c|}{\textbf{\begin{tabular}[c]{@{}c@{}}\rotatebox{90}{Merged By Other User}\end{tabular}}} & 
\multicolumn{1}{c|}{\textbf{\begin{tabular}[c]{@{}c@{}}\rotatebox{90}{\# Review Comments}\end{tabular}}} & 
\multicolumn{1}{c|}{\textbf{\begin{tabular}[c]{@{}c@{}}\rotatebox{90}{\# Comments}\end{tabular}}} & 
\multicolumn{1}{c}{\textbf{\begin{tabular}[c]{@{}c@{}}\rotatebox{90}{\# Lines Changed}\end{tabular}}} \\ 
\hline
\endhead
\hline
\endlastfoot
\multirow{5}{*}{\begin{tabular}[c]{@{}l@{}}kiegroup/\\ kogito-\\runtimes\end{tabular}} & 
\multirow{5}{*}{57.41} & 
\multirow{5}{*}{Maximum} & 
1473 & Yes & 2 & Yes & 7 & 9  & 8  \\ & & & 
1482 & No & 3 & Yes & 0 & 3 & 69 \\ & & & 
1465 & No & 2 & Yes & 0 & 17 & 51 \\ & & & 
1488 & No & 2 & Yes & 0 & 3 & 2 \\ & & & 
1476 & Yes & 4 & Yes & 31 & 25 & 1094 \\ \hline
\multirow{20}{*}{\begin{tabular}[c]{@{}l@{}}apache/\\ activemq-\\artemis\end{tabular}} & 
\multirow{20}{*}{2.44} & 
\multirow{20}{*}{80th} & 
3679 & No & 1 & Yes & 0 & 0 & 6 \\ & & & 
3678 & No & 0 & No & 0 & 0 & 2 \\ & & &
3677 & No & 1 & Yes & 3 & 3 & 74 \\ & & &
3676 & Yes & 3 & Yes & 0 & 15 & 76 \\ & & &
3674 & No & 1 & Yes & 0 & 0 & 2 \\ & & & 
3673 & Yes & 2 & Yes & 15 & 11 & 163 \\ & & & 
3668 & No & 1 & Yes & 0 & 1 & 17 \\ & & &
3665 & Yes & 2 & Yes & 3 & 0 & 67 \\ & & & 
3664 & No & 1 & Yes & 0 & 0 & 9 \\ & & & 
3661 & No & 1 & Yes & 0 & 1 & 57 \\ & & &
3660 & No & 0 & No & 0 & 0 & 4 \\ & & &
3658 & No & 0 & No & 0 & 0 & 828 \\ & & &
3654 & No & 1 & No & 0 & 2 & 235 \\ & & & 
3651 & Yes & 1 & No & 0 & 2 & 698 \\ & & &
3650 & No & 2 & No & 0 & 13 & 5 \\ & & & 
3649 & Yes & 1 & Yes & 11 & 0 & 67 \\ & & & 
3648 & No & 1 & Yes & 0 & 0 & 2 \\ & & &
3647 & No & 0 & No & 0 & 0 & 145 \\ & & &
3645 & No & 1 & Yes & 0 & 0 & 10 \\ & & & 
3641 & No & 0 & No & 0 & 0 & 323 \\ \hline
\multirow{20}{*}{\begin{tabular}[c]{@{}l@{}}radargun/\\ radargun\end{tabular}} & 
\multirow{20}{*}{1.66} & 
\multirow{20}{*}{60th} & 
699 & No & 0 & Yes & 0 & 0 & 6 \\ & & & 
698 & Yes & 1 & Yes & 1 & 1 & 116 \\ & & & 
697 & No & 0 & No & 0 & 0 & 2 \\ & & &
696 & No & 1 & Yes & 0 & 1 & 268 \\ & & &
695 & No & 1 & Yes & 0 & 1 & 10 \\ & & & 
694 & No & 1 & Yes & 0 & 1 & 200 \\ & & & 
692 & No & 2 & Yes & 1 & 0 & 4 \\ & & & 
691 & Yes & 1 & Yes & 2 & 0 & 94 \\ & & &
690 & No & 1 & Yes & 1 & 1 & 296 \\ & & &
689 & Yes & 1 & Yes & 4 & 1 & 95 \\ & & &
688 & No & 1 & Yes & 0 & 1 & 329 \\ & & &
687 & No & 1 & Yes & 0 & 1 & 8 \\ & & &
686 & No & 1 & Yes & 0 & 1 & 2 \\ & & & 
685 & No & 1 & Yes & 1 & 1 & 595 \\ & & &
684 & No & 1 & Yes & 0 & 0 & 27 \\ & & &
683 & No & 0 & No & 0 & 0 & 2 \\ & & &
682 & No & 1 & Yes & 0 & 0 & 42 \\ & & &
681 & No & 1 & Yes & 0 & 0 & 4 \\ & & &
680 & No & 1 & Yes & 0 & 1 & 11 \\ & & & 
678 & No & 0 & No & 0 & 0 & 2 \\ \hline
\multirow{20}{*}{\begin{tabular}[c]{@{}l@{}}wocommunity/\\ wonder\end{tabular}} & 
\multirow{20}{*}{0.54} & 
\multirow{20}{*}{40th} & 
954 & No & 1 & Yes & 0 & 3 & 3 \\ & & & 
953 & No & 1 & Yes & 0 & 0 & 2 \\ & & & 
950 & No & 1 & Yes & 0 & 0 & 4 \\ & & & 
949 & No & 1 & Yes & 0 & 1 & 5 \\ & & & 
948 & No & 1 & Yes & 0 & 0 & 2 \\ & & & 
946 & No & 1 & Yes & 0 & 0 & 14 \\ & & & 
945 & No & 1 & Yes & 0 & 3 & 6 \\ & & & 
940 & No & 1 & Yes & 0 & 0 & 2 \\ & & & 
938 & No & 1 & Yes & 0 & 0 & 4 \\ & & & 
937 & No & 2 & Yes & 0 & 1 & 9 \\ & & & 
936 & No & 1 & Yes & 0 & 0 & 2 \\ & & & 
934 & No & 1 & Yes & 0 & 0 & 3 \\ & & & 
933 & Yes & 2 & Yes & 0 & 3 & 978 \\ & & & 
926 & No & 2 & Yes & 0 & 2 & 16 \\ & & & 
925 & No & 1 & Yes & 0 & 3 & 4 \\ & & & 
924 & No & 1 & Yes & 0 & 0 & 3 \\ & & & 
923 & No & 1 & Yes & 0 & 0 & 13 \\ & & & 
920 & No & 1 & Yes & 0 & 0 & 77 \\ & & & 
919 & No & 1 & Yes & 0 & 0 & 56 \\ & & & 
918 & No & 1 & Yes & 0 & 1 & 2 \\ \hline
\multirow{5}{*}{\begin{tabular}[c]{@{}l@{}}uwetrottmann/\\ seriesguide\end{tabular}} & 
\multirow{5}{*}{0.15} & 
\multirow{5}{*}{20th} & 
787 & No & 1 & Yes & 0 & 1 & 6 \\ & & & 
798 & No & 0 & No & 0 & 1 & 584 \\ & & & 
769 & No & 0 & No & 0 & 0 & 45 \\ & & & 
810 & No & 0 & No & 0 & 1 & 11456 \\ & & & 
790 & No & 0 & No & 0 & 1 & 986 \\ \hline
\multirow{5}{*}{\begin{tabular}[c]{@{}l@{}}komamitsu/\\ luency\end{tabular}} & 
\multirow{5}{*}{0} & 
\multirow{5}{*}{Minimum} & 
199 & No & 0 & No & 0 & 0 & 2 \\ & & & 
223 & No & 0 & No & 0 & 0 & 3 \\ & & & 
192 & No & 0 & No & 0 & 0 & 2 \\ & & & 
248 & No & 0 & No & 0 & 0 & 2 \\ & & & 
204 & No & 0 & No & 0 & 0 & 2 \\ 
\bottomrule
\caption{\label{tab:code-review-validation}\revised{Summary of code review information in a number of selected projects based on their code review intensity. The code review intensity was calculated over the 20 PRs that we have analyzed per project.}}
\end{longtable}

\revised{
For the validation, we consider the following six observations with regard to code reviews. 
The fifth column of Table~\ref{tab:code-review-validation}, entitled \textbf{Change Requested}, indicates whether any change is requested by 
the reviewer(s) during the discussion and whether there is a commit after that request to apply the requested change. The sixth column 
indicates the number of reviewers involved (\textbf{\# Reviewers}), and \textbf{Merged By Other User} indicates whether the pull request 
is merged by any user other than the creator/author of the pull request. Then, in column 8 we list \textbf{\# Review Comments}, 
specifically the comments on a portion of the unified diff made during a pull request review. Column 9 shows the \textbf{\# Comments} (comments on discussions part, except the first one and the ones by bots), and column 10 tabulates the \textbf{\# Lines Changed} (in all the commits of a pull request). 
}

\revised{
Table~\ref{tab:code-review-validation} depicts a number of projects and their associated code review metric values. We focus on the minimum, maximum, 
and the code review intensity values at the 20th, 40th, 60th, 80th percentiles, to look at a good spread in terms of code review intensity values. Accordingly, 
we investigate whether if the code review intensity value goes up, the other indicates are typically also increasing. 
As an example, the project with the highest code review intensity (\textit{kiegroup/kogito-runtimes}) has the highest number of reviewers per pull request on 
average. Moreover, regardless of the amount of changes, it has at least 2 reviewers for each pull request. 
Whereas, if we look at the project with the lowest code review intensity (\textit{komamitsu/luency}), we see 
that there is no sign of having any kind of code review before merging the pull requests. A potential explanation can be that they use another platform 
to do their code reviews (as mentioned in our threats to validity), but for this specific project this seems unlikely, as there is only a single contributor. 
Also, by looking at the project at the 20th percentile (\textit{uwetrottmann/seriesguide}), 
we see for 4 pull requests out of 5 that the user creating the pull requests is also merging them. In addition, we see very small numbers of comments being made. 
}
\revised{For projects at the 40th, 60th, and 80th percentile, we obtain very similar observations in terms of number of reviewers, number of comments, etc. in 
the 5 randomly selected pull requests for this analysis. That is why for these categories, we expanded our investigation to 20 pull requests in total. Looking at 
the number of reviewers, or whether a pull request was merged by another user, we can not make a definite judgement about the 
code review intensity, since the differences in observations is not obvious. But, if we look at the change requested column 
we see that for projects in 80th, 60th, and 40th percentile in total have respectively 5, 3, and 1 changes requested, which 
follows our code review intensity metric values. Moreover, 
the total number of review comments for projects in 80th, 60th, and 40th percentile are respectively 32, 10, and 0.
And the total number of their discussions comments are 48, 11, and 17. Adding up both numbers gives us 
80, 21 and 17 review comments \& discussion comments in their last 20 pull requests, showing that the intensity of code 
review decreases as we go from the 80th percentile to the 40th percentile category. 
}

\subsubsection{\textbf{Step 5. Testing}} \label{testing-data}
To establish whether software projects do testing and at what level of intensity, we measure branch coverage. In order to do so, 
in this step we build the projects (that we have already successfully built in Step 1) and run their tests. We did remove some buildable projects from our dataset, in case those projects had no code or no tests. For all build configuration files in a project (\texttt{build.gradle} for Gradle or \texttt{pom.xml} for Maven) we search for the \texttt{src/main} path for source code and the \texttt{src/test} path for tests in the same directory. If we could not find any code or test directory for that project then we marked them as projects with no code or no test and removed them for this analysis step. 

To establish code coverage, we add the JaCoCo plugin\footnote{JaCoCo \url{https://www.jacoco.org}, last visited May 20th 2022.} to the projects' build configuration file. 
We choose JaCoCo because of its popularity, slightly higher visibility, and easier integration~\cite{DBLP:journals/sqj/HorvathGBTBG19}. 
However, integrating JaCoCo with Gradle or Maven projects is not a trivial task. To integrate JaCoCo with a project we need to insert specific configurations in projects' 
build file (\texttt{pom.xml} or \texttt{build.gradle}). Depending on the structure of the project (having sub-projects or parent-projects), usage of other plugins, build 
system of the project (Gradle or Maven), and other configurations (in some cases tests were disabled manually in build configuration) we had to change specific lines of 
the build configuration. Moreover, for Gradle projects there are two different domain specific language (DSL), i.e., Groovy and Kotlin which makes it non-trivial to automatically 
apply the JaCoCo plugin to projects without causing a build failure. 

The process of automatically integrating the JaCoCo plugin for hundreds of projects has proven extremely cumbersome. To the best of our knowledge, we are the first to have created an automatic process for this task. To establish an automated process that works for a majority of projects, we had to finetune our script to cater for different build script structures. 
Moreover, for Maven projects we usually had to build projects in several steps, since it is quite common for a project to have multiple \texttt{pom.xml} files, e.g., a \texttt{pom.xml} file per module. 
On the other hand, for Gradle projects we found it easier to build a project, since when building the parent directory all child modules would be built. In the case of Maven projects, we have encountered projects were we could not produce complete coverage results, because not all modules of a project could be built successfully. Also, there are other reasons for not having complete coverage results for a project, for example, when not all modules of a project have tests, when  unconventional paths for source code and test code folders are used, or when our general approach of automatically inserting the JaCoCo commands into the build file conflict with specific build file configurations of a project. For this special configurations, we would need a manual action to insert the JaCoCo command. This is one reason why we have partial coverage results for some projects. 

\revised{
\subsubsection{\textbf{Mature Projects}}\label{mature-projects-data}}
\revised{
In the results section, we will investigate how the maturity of projects possibly affects the 
results of our analysis. For each of the analysis steps that we perform, we will compare how a 
practice is either followed similarly or differently among mature projects compared to the overall set 
of projects.
}

\revised{
We first need to establish what we consider to be a mature project.
We specifically focus on two indicators of projects, namely \emph{popularity} and \emph{activity}. 
As a proxy for the activity of a project, we consider the following project attributes: the count of commits, 
pull requests, issues, and contributors. Projects that we earmark as being more active, have higher counts
of the aforementioned attributes. 
To approximate the popularity of a project, we take the number of forks, starts, and watchers into consideration, 
again considering higher values for these attributes as an indication that a project is more popular. 
We consider a mature project as a project that has higher activity and higher popularity.}

\revised{
We have used the following procedure to select mature projects in our dataset. 
For each of the aforementioned attributes, we first select the projects that have a score for that attribute that is in the top 10\%.
Since the distribution of the attributes is right skewed, projects that are at the 90th percentile and up have (very) high values for a
particular attribute, making it reasonable to consider them as active or popular. However, as an extra condition for establishing 
that these projects are indeed active or popular, we require them to be at the 90th percentile or above for at least two popularity 
attributes, \emph{and} two activity attributes.
}

\section{Results} \label{section:results}
In this section we will discuss the results of each of the analysis steps that are depicted in Figure~\ref{fig:data-collection-overview}, 
which represents our data processing workflow. 
After presenting the results per analysis step, 
we elaborate on how the different quality assurance practices are used in parallel in Section~\ref{section:put-together}. 

\revised{
In order to investigate whether mature projects are implementing the quality assurance practices in the same way as other 
projects, we follow the procedure of Section~\ref{mature-projects-data}. In doing so, we have established 50 out of 1454 
projects as being more mature than most projects. Our subsequent analyses of mature projects is thus based on these 
50 projects\footnote{List of these projects and the details on the analysis are available in our replication package.}.
}

\revised{
The results presented in each part of this section will be used in our discussion in Section~\ref{section:discussion} 
to present a global answer to our research questions. 
}

\subsection{Local Builds}\label{section:localbuilds}
As explained in Section~\ref{local-builds-data}, starting from a set of 1454 projects in our study, we filtered out non-Gradle, non-Maven and Android projects. Moreover, we also removed Gradle and Maven Projects that did not have any build configuration file in their root directory. Furthermore, in cases where we recognized both Gradle and Maven configuration files inside a project, we have labelled them as Gradle projects, our reasoning being that the Maven configuration file might be a legacy element, as the Gradle ($^{\circ}$2008) build system is newer than Maven ($^{\circ}$2004). The aforementioned steps left us with a total of 1047 projects that we could build locally ($\sim$72\% of the original dataset). 

\begin{figure}
    \centering
    \includegraphics[width=100mm]{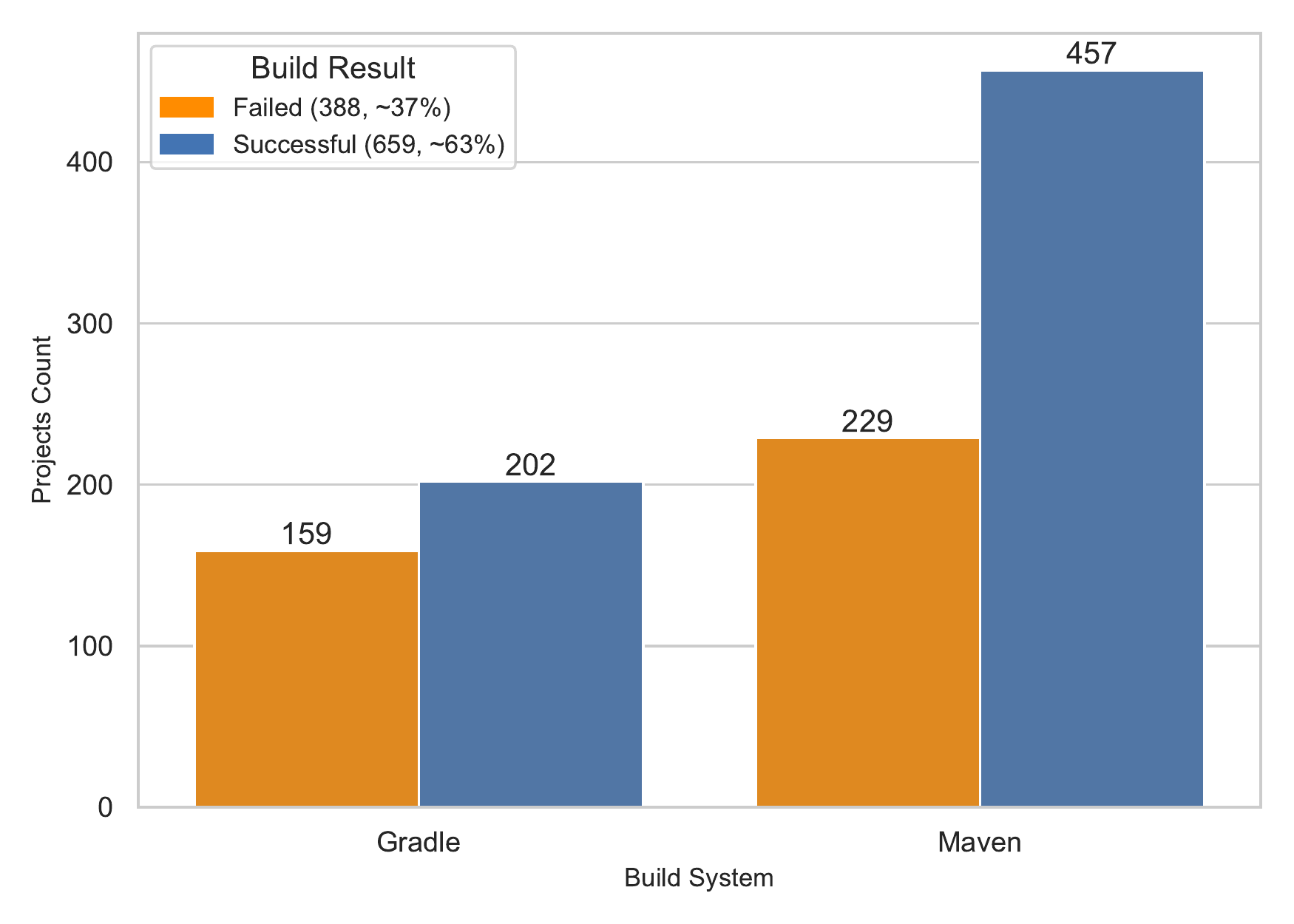}
    \caption{Summary of local build results based on build systems.}
    \label{fig:local-builds-bar-chart}
\end{figure}

To filter successful and failed builds we searched for BUILD SUCCESSFUL / BUILD SUCCESS and BUILD FAILED / BUILD FAILURE phrases in respectively Gradle and Maven build logs. However, in some cases none of the aforementioned phrases were found in build logs due to: (for Gradle projects) missing libraries that prevented running the build process, encountering errors when processing Maven's \texttt{pom.xml} (Project Object Model) file, and encountering errors because of missing environment variables configuration. Even though there was no explicit build failure phrase in the build log, we labeled these builds as failed since no successful build was achieved. Moreover, we also consider the following cases as failed builds: builds in which the build command could not be executed (e.g., due to a missing commit in git tree), and builds that could not finish their execution (reaching the one hour time limit). 
The summary of local build results is shown in Figure~\ref{fig:local-builds-bar-chart}.
From this graph, we observe that $\sim$63\% of projects could be built successfully out of the box. 
We do observe that that despite Gradle being a newer and more popular build system~\cite{DBLP:journals/ese/McIntoshNAMH15}, among our selection of projects, it exhibits a  relatively lower successful build percentage (56\%) compared to Maven (66.6\%).

We analysed a selection of failed builds to better understand why builds fail. To this end, we first randomly selected 50 projects from the total of 388 projects with build failures, then, we manually labeled them according to their build log outputs. In Table~\ref{tab:build-failure-reasons} we list the most frequently found build failure categories from our sample. A key observation that we can make is that build failures related to issues with dependencies and compilation errors account for around half of build failures. For 11 projects the cause of failure was unclear due to limited build logs, which were excluded.

\begin{table}[!t]
\centering
\begin{smaller}
\begin{tabular}{l|l|l}
\toprule
    Category & \# & \%    \\ 
    \midrule
    Compilation          & 13 & 26\%  \\
    Dependency           & 10 & 20\%  \\
    Configuration        & 6  & 12\%  \\
    Parsing                & 5  & 10\%  \\
    Assembling             & 3  & 6\%  \\
    Other			& 2 & 4\% \\
\bottomrule
\end{tabular}
\end{smaller}
\caption{\label{tab:build-failure-reasons}Top categories of local build failure errors}
\end{table}

\begin{center}
    \fbox{\begin{minipage}{0.9\columnwidth}
        \textbf{Observation 1:} $\sim$63\% of projects build successfully and Maven projects show a higher percentage of successful builds ($\sim$67\%) compared to Gradle projects (56\%).
    \end{minipage}}
\end{center}

\revised{\textbf{\subsubsection{Local Builds in Mature Projects}}}
\revised{Looking at the local build results among the 50 mature projects, we see that 21 projects build successfully out of the box, 
from the total number of 33 eligible projects to build locally. Meaning, 64\% of mature projects build locally, which is nearly 
the same ratio when comparing to other projects in our dataset.}

\revised{
    \begin{center}
        \fbox{\begin{minipage}{0.9\columnwidth}
            \textbf{Observation 1*:} Mature projects have roughly the same percentage of local build success (64\%) comparing to all the 
            projects in our dataset ($\sim$63\%).
        \end{minipage}}
    \end{center}
}

\subsection{Automatic Static Analysis Tools}
Overall, we observe that in 38\% (548) of all 1454 projects in our study we found evidence of Checkstyle, Findbugs, or SpotBugs Automatic Static Analysis Tools (ASATs). Table~\ref{tab:3} shows the frequency of ASAT usage, and also indicates that no project uses more than 2 ASATs in parallel. Among our set of studied projects, Checkstyle is the most popular ASAT with $\sim$36\% projects using it. 

\begin{center}
    \fbox{\begin{minipage}{0.9\columnwidth}
        \textbf{Observation 2:} 38\% of projects use at least one ASAT.
    \end{minipage}}
\end{center}

\revised{\textbf{\subsubsection{ASATs Usage Among Mature Projects}}}
\revised{Unlike local builds, we observe a great difference in the ratio of 
mature projects using ASATs. We see evidence of ASATs usage in 70\% 
of mature projects (35 out of all 50). This shows mature projects tend to
invest more in ASATs. 
}

\revised{
    \begin{center}
        \fbox{\begin{minipage}{0.9\columnwidth}
            \textbf{Observation 2*:} 70\% of mature projects use at least one ASAT.
        \end{minipage}}
    \end{center}
}

\begin{table}[!t]
\centering
\begin{smaller}
\renewcommand{\tabcolsep}{0.8mm}
\begin{tabular}{c|ccc|ccc}
\toprule
\begin{tabular}[c]{@{}c@{}} All\\  Projects\end{tabular} & \begin{tabular}[c]{@{}c@{}}Use\\ CheckStyle\end{tabular} & \begin{tabular}[c]{@{}c@{}}Use\\ FindBugs\end{tabular} & \begin{tabular}[c]{@{}c@{}}Use\\ SpotBugs\end{tabular} & \begin{tabular}[c]{@{}c@{}}Use\\ no ASAT\end{tabular} &\begin{tabular}[c]{@{}c@{}}Use\\ 1 ASAT\end{tabular} & \begin{tabular}[c]{@{}c@{}}Use\\ 2 ASATs\end{tabular} \\ \midrule
    1454                                                    & 519 ($\sim$36\%)                                         & 3 ($\sim$0\%)                                          & 104 ($\sim$7\%)                                        & 906 ($\sim$62\%) &470 ($\sim$32\%)                                     & 78 ($\sim$5\%)       \\
    \bottomrule                                
\end{tabular}
\end{smaller}
\caption{\label{tab:3}ASATs usage results.}
\end{table}

\subsection{Continuous Integration} \label{ci-results}
We set out to study two aspects of Continuous Integration: firstly, the prevalence of the use of CI in our study set, and secondly, a comparison between CI build results and local build results. As stated in Section~\ref{ci-data}, data about CI related activities are derived from GitHub status checks. We will study the usage of status checks separately.

\subsubsection{\textbf{Prevalence of CI}}
As explained in Section~\ref{ci-data}, we study the use of CI for a project at a specific point in time, namely the latest commit hash in our dataset. At that point, we have results stemming from 1131 projects using external services that mark their \textit{Commit Status} or are using GitHub Actions to run \textit{Checks} in the form of workflows. Overall, we found that checks (6,906 checks) are more frequently used by projects compared to statuses (1,224 statuses). However, not all checks and statuses can be considered as a use of CI. So, we came up with a heuristic to filter CI related ones (explained in Section~\ref{ci-data}). After this filtering we established that 912 (62.7\%) projects use status checks in their CI pipelines: 613 projects only used CI checks, 200 projects only used CI commit statuses, and 99 projects used both (more details in Table~\ref{tab:ci-usage}).

\begin{table}[]
\centering
\begin{smaller}
\begin{tabular}{l|l|llll}
\toprule
\# All Projects 	& \# Projects & \multicolumn{3}{l}{\# Projects} \\
			& \emph{Not} Using CI & \multicolumn{3}{l}{Using CI} \\
\midrule
		&				& 			& Use Checks	& 613 \\
	 	&	 			&  			& Use Statuses	& 200 \\
1454		&	542 (37\%)	& 912 (63\%)	& Use Both	& 99 \\ \cline{4-5}
		&				&			& Successful State & 695 (76.2\%) \\
		&				&			& Unsuccessful State & 219 (23.8\%) \\
\bottomrule
\end{tabular}
\end{smaller}
\caption{\label{tab:ci-usage}Counts of projects using CI and the rollup result of CI status checks. Projects might be using checks or commit statuses (or both), either way they are considered as projects using/following CI practice.}
\end{table}

Looking at the rollup state of status checks of projects using CI, we obtained an overview of CI results of our projects. In doing so, we observe that 695 (76.2\%) of the 912 projects using CI have a successful state. The other 219 (23.8\%) projects had an unsuccessful state (2 pending and 217 failure).
 
\begin{figure}
    \centering
    \includegraphics[width=70mm]{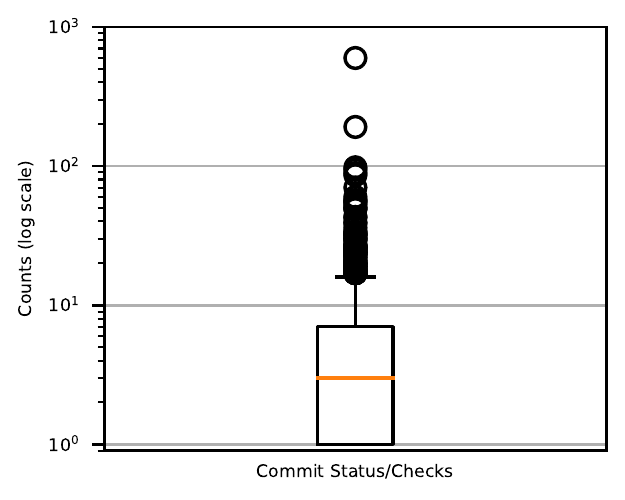}
    \caption{Counts of status checks among projects following CI practice.}
    \label{fig:checks-statuses}
\end{figure}

Moreover, when we look at counts of status checks on the latest commit of projects, we observe that projects do not have the same approach when following CI practice: there are projects with more than 100 $\geq$ CI checks and also projects with only one CI check. Figure~\ref{fig:checks-statuses} depicts the frequency of the number of status checks in projects using CI (median = 3, and average = 7.35). 

To have an idea of why some of the projects have a high number of status checks and what is the purpose of it, we selected one project with more than 50 status checks, namely the Checkstyle project which has 60 status checks\footnote{\url{https://github.com/checkstyle/checkstyle/search?q=4ce315d27d387b48befd667c7ed8426a1bad6dbb&type=commits}}. The project runs many GitHub Apps (such as GitHub Actions App, Cirrus CI, Travis CI, Azure Pipelines, and CloudBees). Additionally, Checkstyle uses many external services to mark a commit's status (such as Drone CI, Semaphore CI, AppVeyor, Circle CI, and Wercker). When analysing these status checks, we observed some of the motives for running them, including: checking broken links / closed issue references, code analysis to discover vulnerabilities across the codebase using CodeQL, running mutation analysis for different modules using PIT, and building the project with different versions of the JDK and OS (so-called multi-environment builds, see Beller et al.~\cite{DBLP:conf/msr/BellerGZ17}). Outside the scope of our current investigation, it would be interesting to study the various checks that software projects use in more detail in future work to (1) better understand what quality aspects developers care about, and (2) why several CI services and GitHub apps are used in conjunction. 

\begin{figure}
    \centering
    \includegraphics[width=100mm]{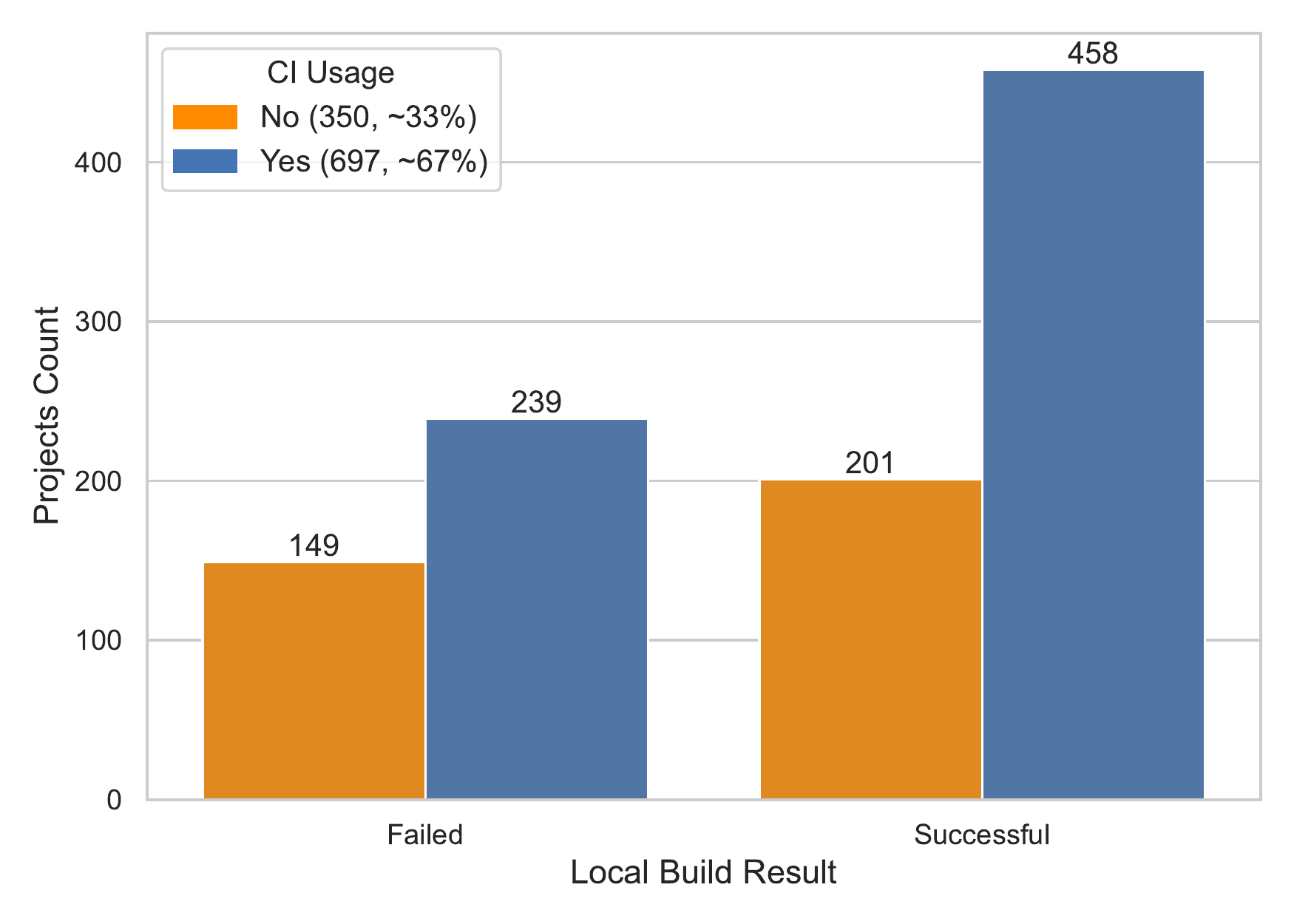}
    \caption{Summary of CI usage in projects based on their local build results.}
    \label{fig:ci-usage-bar-chart}
\end{figure}

With the CI usage results at hand in Table \ref{tab:ci-usage}, and with our results from local builds available in Section~\ref{section:localbuilds}, we set out to investigate the relation between using CI and buildability. Specifically, we aim to understand whether projects that use CI in their development process are more likely to build locally without failure? Figure~\ref{fig:ci-usage-bar-chart} shows the summary of CI usage based on our local build results. Using the Chi-square test ($\chi ^2 \simeq 6.5$, p-value $\simeq$ 0.01 $<$ 0.05, and Cramer's V $\simeq$ 0.08), there is evidence that a weak, yet statistically significant correlation between CI usage and building locally exists\footnote{The effect size interpretation for Cramer's V is: 0 $\leq$ V $\leq$ 0.2 is weak, 0.2 $<$ V $\leq$ 0.6 is moderate, V $>$ 0.6 is strong}. 

\begin{center}
    \fbox{\begin{minipage}{0.9\columnwidth}
        \textbf{Observation 3:} $\sim$63\% of projects use CI with $\sim$76\% of them having a successful state. 
        There is a weak, yet statistically significant positive correlation between CI usage and building successfully locally, meaning that using CI does increase the likelihood of a successful local build.
    \end{minipage}}
\end{center}

\subsubsection{\textbf{CI Builds vs. Local Builds}}
In Section~\ref{section:localbuilds} we have reported on the local build results of projects. Now with the CI results available, we select CI-related checks from the entire collection of checks to compare them with our local builds. The goal of this comparison is to see whether having a successful/failed build on CI is correlated with a successful/failed build on a local machine. Furthermore, we will dive in to see if there is any relation between having more CI builds and a successful build on a local machine. Our reasoning here is to see if building several times on CI, which we assume to be multi-environment builds, can increase projects' buildability on a local machine with common settings.

After filtering build related CI status checks (explained in Section \ref{ci-data}) 
we obtain 2846 build status checks for a total of 843 projects. It is obvious that there exists more than one CI build for some projects, thus we looked into the data to see what are the reasons: building with different versions of JDK / operating systems (e.g., multiple integration environments~\cite{DBLP:conf/msr/BellerGZ17}), building different modules separately, building and running tests in different blocks, and building with/without cleaning cache.

\begin{figure}
    \centering
    \includegraphics[width=100mm]{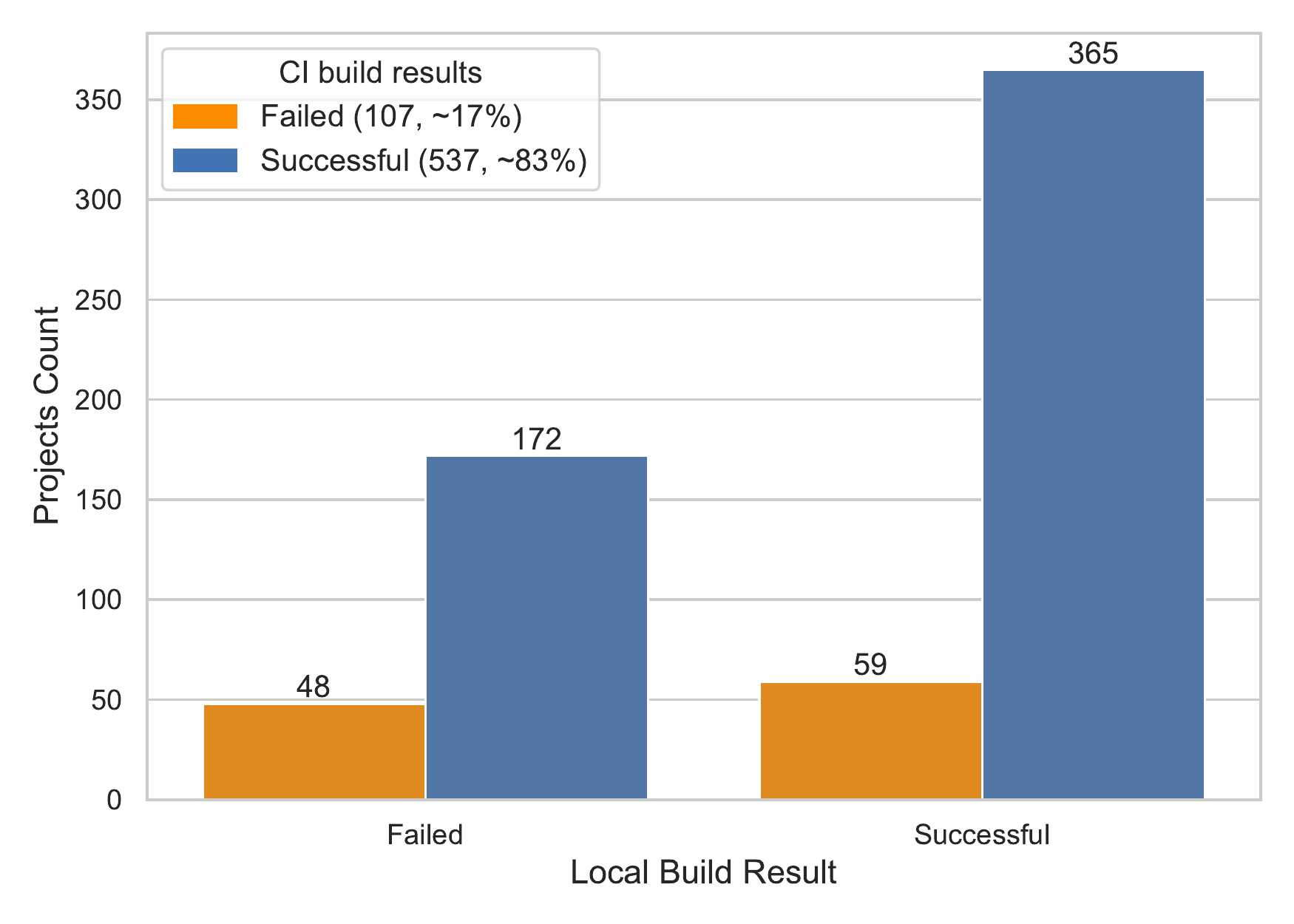}
    \caption{Summary of CI build results based on local builds.}
    \label{fig:ci-local-builds-bar-chart}
\end{figure}

Next, we have combined CI build results for each of the projects to then compare the overall CI build result to the build results of the local machine (explained in Section~\ref{ci-data}). Overall, we observe 83.5\% successful CI build results among 843 projects using CI build status checks. 
Next, we have compared CI build results with local build results using the Chi-square test. We have obtained values $\chi ^2 \simeq 6$, p-value $\simeq$ 0.014 $<$ 0.05, and Cramer's V $\simeq$ 0.10. This indicates a weak, yet statistically significant correlation.
To put things into context, in 64\% of projects (from a total of 644 that we had both CI/local build results) we found the same CI and local build results (shown in Figure~\ref{fig:ci-local-builds-bar-chart}).

\begin{center}
    \fbox{\begin{minipage}{0.9\columnwidth}
        \textbf{Observation 4:} For $\sim$64\% of projects we found matching CI build related status checks and local build outcomes. Using the Chi-square test, we observe a statistically significant weak correlation between successful local builds and CI builds.
          \end{minipage}}
\end{center}

\revised{\subsubsection{\textbf{Prevalence of CI in Mature Projects}}}
\revised{
The CI usage ratio among mature projects is almost the same as what we observe in 
all the projects in our dataset. 62\% of mature projects are using CI (using the same 
heuristic method to filter CI usage). Also, among those using CI, 16 (out of 31) projects 
had a successful rollup state. This is $\sim$52\%, which 
is less than the $\sim$76\% successful CI state among all the projects. 
One question worth raising here is why mature projects have less successful CI states? 
A preliminary investigate shows that mature projects have more status checks in their CI workflows. 
Specifically, on average they have 10 (median = 4) status checks, compared to other projects that have on average 7 (median = 3) checks. 
However, the number of projects in our mature set is not big enough to show statistical significance.}

\revised{
    \begin{center}
        \fbox{\begin{minipage}{0.9\columnwidth}
            \textbf{Observation 4*:} We observe that 62\% of mature projects use CI with $\sim$52\% of them having a 
            successful state. When we compare mature projects against the general set of projects, we see that a similar
            ratio of projects use CI, however mature projects have a lower percentage of successful CI states, $\sim$52\%
            compared to $\sim$76\%.      
       	\end{minipage}}
    \end{center}
}

\subsection{Code Review}\label{section:results:codereview}
\revised{As a quality assurance practice code review has been shown to be able to improve 
the quality of software, quality of changes, and quality of system design~\cite{DBLP:conf/wcre/ThongtanunamMHI18}.} 
To study how code reviewing is done in our set of GitHub projects, we investigate the users' activities that are related to code reviewing during the merging of pull requests (PRs) (also see Section~\ref{code-review-data}).
More specifically, we collect the count of comments, reviews, total changed lines, and total changed files for the last 20 merged PRs of 1398 projects. Another 56 projects did not meet our data collection condition of having at least 20 merged PRs while checking the last 100 PRs.

\begin{figure}[!t]
    \centering
    \includegraphics[width=100mm]{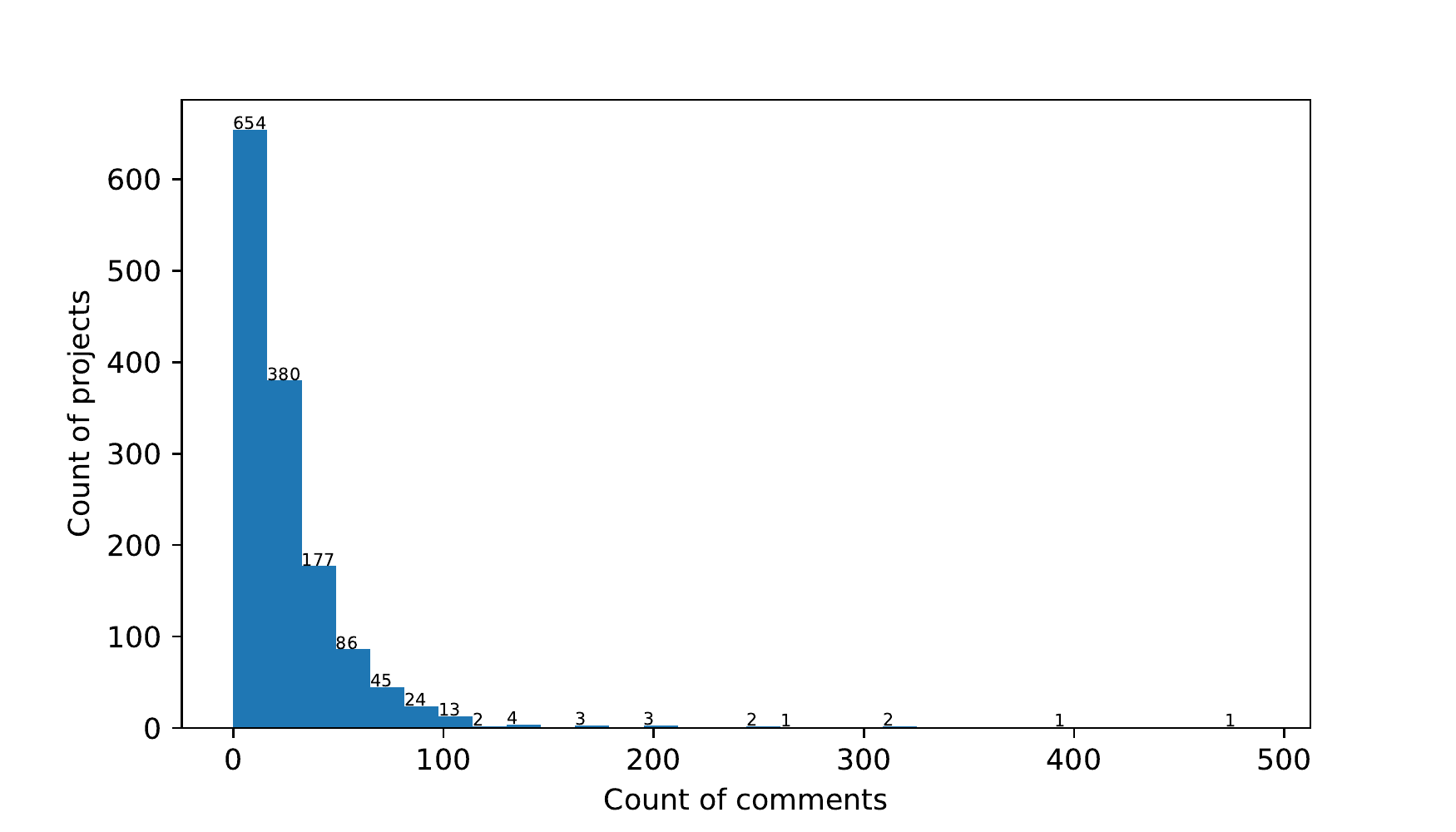}
    \caption{Distribution of count of comments in projects' last 20 merged PRs.}
    \label{fig:code-review-comments-bar-chart}
\end{figure}

\begin{figure}[!t]
    \centering
    \includegraphics[width=100mm]{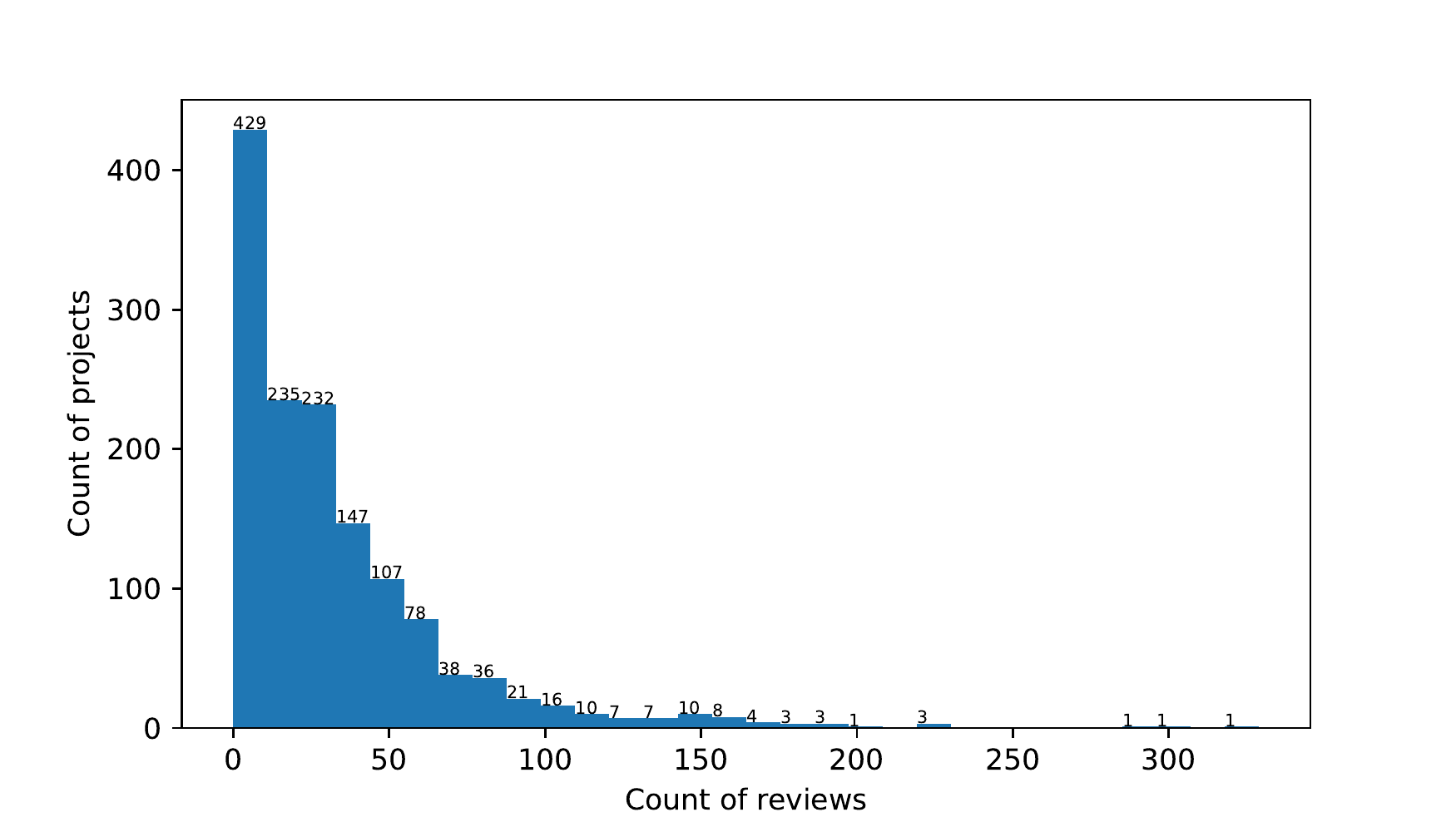}
    \caption{Distribution of count of reviews in projects' last 20 merged PRs.}
    \label{fig:code-review-reviews-bar-chart}
\end{figure}

To study the state of code review as a quality assurance practice, we start without making any assumptions with regard to what can be expected in terms of the intensity of code review activities. 
To keep our assumptions in check, we use a benchmarking approach that is taken from Alves \etal~\cite{DBLP:conf/icsm/AlvesYV10}. Specifically, the benchmarking enables us to categorise 
projects into four different groups: projects with very high, high, moderate, and low code reviewing. To determine the thresholds for the 4 categories we use the 70th, 80th, and 90th percentiles of 
the sum of count of reviews and comments in each of projects as the boundaries. These specific boundaries have previously been empirically established and used in the SIG quality model from 
Alves \etal~\cite{DBLP:conf/icsm/AlvesYV10} and the test code quality model of Athanasiou \etal~\cite{DBLP:journals/tse/AthanasiouNVZ14}. An important prerequisite to being able to use this
benchmarking approach is that the distribution of values follows a power-law distribution. In order to determine whether the distribution of count of reviews and comments indeed follows a power-law,
we plot the values for our metrics for the last 20 merged PRs in the projects that we consider in Figure~\ref{fig:code-review-comments-bar-chart} and~\ref{fig:code-review-reviews-bar-chart}. 
Both metrics seemingly follow a power-law-like distribution: count of reviews and comments are mostly low, but there are a few cases with higher values. As such, we can apply Alves \etal's calibration method to derive different categories of code reviewing intensity~\cite{DBLP:conf/icsm/AlvesYV10}.

Considering the last 20 merged PRs, projects with less than 38 reviews fall in the low level of code reviewing, projects with review counts between 38 and 50 are classified as doing a moderate level of code reviewing, the ones with review counts between 50 and 73 are considered as projects with high levels of code reviewing, and a review count higher than 73 makes us consider the project doing a very high level of code reviewing. Similar, for comments less than 29, between 29 and 38, between 38 and 56.3, and greater than 56.3 we have low, moderate, high, and very high level of code reviewing. The mentioned thresholds corresponding to percentiles are shown in Table~\ref{tab:7}.


\begin{table}[!t]
\centering
\begin{smaller}
    \begin{tabular}{l|cccc}
    \toprule
    \textbf{Metric} & \textbf{\begin{tabular}[c]{@{}c@{}}Low\\ Level\end{tabular}} & \textbf{\begin{tabular}[c]{@{}c@{}}Moderate\\ Level\end{tabular}} & \textbf{\begin{tabular}[c]{@{}c@{}}High\\ Level\end{tabular}} & \textbf{\begin{tabular}[c]{@{}c@{}}Very High\\ Level\end{tabular}} \\ \midrule
    Reviews Count   & 38                                                           & 50                                                                & 73                                                            & \textgreater{}73                                                   \\ \midrule
    Comments Count  & 29                                                           & 38                                                                & 56.3                                                          & \textgreater{}56.3 \\
    \bottomrule                                                
\end{tabular}
\end{smaller}
\caption{\label{tab:7}Thresholds for code review metrics.}
\end{table}

\begin{center}
    \fbox{\begin{minipage}{0.9\columnwidth}
        \textbf{Observation 5:} Projects having more than 50 reviews or 38 comments in their last 20 merged pull requests are among the projects doing more intense code reviewing. 
        Similarly, projects having less than 38 reviews or 29 comments can be considered as projects doing relatively less code reviews.
    \end{minipage}}
\end{center}

\revised{\subsubsection{\textbf{Code Review in Mature Projects}}}
\revised{
    We observe that 48\% of mature projects have either a high or very high level of code review intensity 
    based on the thresholds measured in Table~\ref{tab:7}. Around 26\% of them fall in the category of \emph{low}
    code review for both reviews and comments count. 
    If we do a comparison with the code reviewing intensity in all the projects, we see that 30\% of 
    projects having high or very high level of interviews and 54.5\% of them falling below the 
    low level. This shows mature projects tend to invest more in their code reviews as a quality 
    assurance practice.
}

\revised{
\begin{center}
    \fbox{\begin{minipage}{0.9\columnwidth}
        \textbf{Observation 5*:} Around 48\% of mature projects have either a high or very high level of code review 
        intensity. Circa 26\% of the projects have a low level of code review intensity.
    \end{minipage}}
\end{center}
}
\subsection{Testing}\label{section:results:testing}
\revised{
Software testing is a practice to improve the quality of software 
by detecting faults in the source code~\cite{DBLP:conf/icse/GopinathJG14}.} 
The goal of this section is to see how well projects are doing in terms of \revised{software} testing. 
In contrast to code reviewing, we have established metrics for measuring the quality of tests, namely code coverage metrics~\cite{Zhu1997}. In addition, to gauge how well software projects are tested in terms of code coverage, we can rely on the existing classification scheme for unit test coverage proposed by Heitlager \etal~\cite{DBLP:conf/quatic/HeitlagerKV07}. This classification ranks projects in 5 different levels according to their unit test coverage and makes it easier to reason about testing (see  Table~\ref{tab:8}). 
In our investigation, we rely on \emph{branch coverage} as calculated by the JaCoCo tool.\footnote{\url{https://www.jacoco.org}, last visited May 20th, 2022.} 

As mentioned in Section~\ref{testing-data} to see the state of testing in our projects, we selected 599 projects that (1) could be built successfully in the first step of our study and (2) had conventional paths for their source code and test code (at least in one of their subprojects / modules). Having conventional paths for source code and tests means that we could detect their path in a project. We also repeated this for all subprojects / modules inside a project before applying JaCoCo. If source code and tests paths were detected, then we would apply JaCoCo plugin and build them to generate coverage results. The strict interpretation of the necessity to have conventional paths for the tests does imply that there might be projects that do have high quality tests, but that are not part of our analysis. 

\begin{figure}[!t]
    \centering
    \includegraphics[width=70mm]{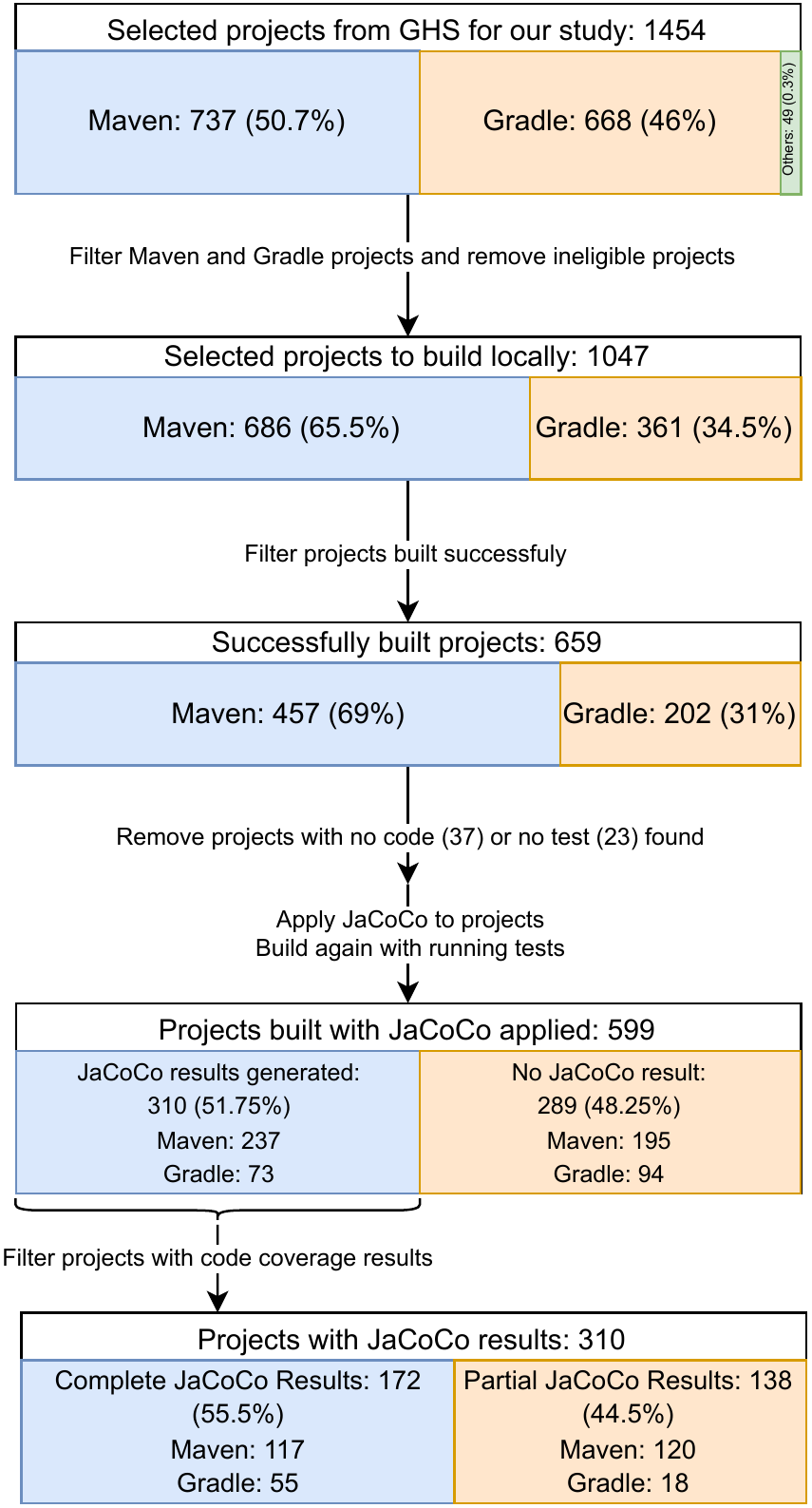}
    \caption{Overview of studied projects in building and testing.}
    \label{fig:testing-results-overview}
\end{figure}

For the 599 projects that we applied JaCoCo to, we obtain JaCoCo test coverage results for 310 of them (51.75\%). From these 310 projects, we observe complete coverage results for  172 (28.7\%) projects, i.e., we did manage to get coverage results for all of their modules. For another 138 (23\%) projects, we could only collect partial coverage results, i.e., we have coverage results for some of the modules. We did not manage to obtain coverage results for the remaining 289 (48.3\%) projects. An overview of the aforementioned process can be witnessed in the bottom part of Figure~\ref{fig:testing-results-overview}.

\medskip

\noindent \textbf{Projects with no coverage results.} To investigate the reasons of 289 projects having no JaCoCo output, we randomly selected 50 of them and manually checked their log files. These projects had at least one module with code \emph{and} tests detected, but we could not see any coverage results after applying the JaCoCo plugin. The most common reasons for this were: (1) JaCoCo failing or skipping its execution due to some (mis)configuration in the project's build setting, or conflicting with other settings of project, and (2) build failure. See Table~\ref{tab:jacoco-failure-reasons} for a global overview of reasons. Moreover, for 12 projects the build logs were very sparse, and we could thus not establish the reason of having no coverage result.

\medskip

\begin{figure}
    \centering
    \includegraphics[width=80mm]{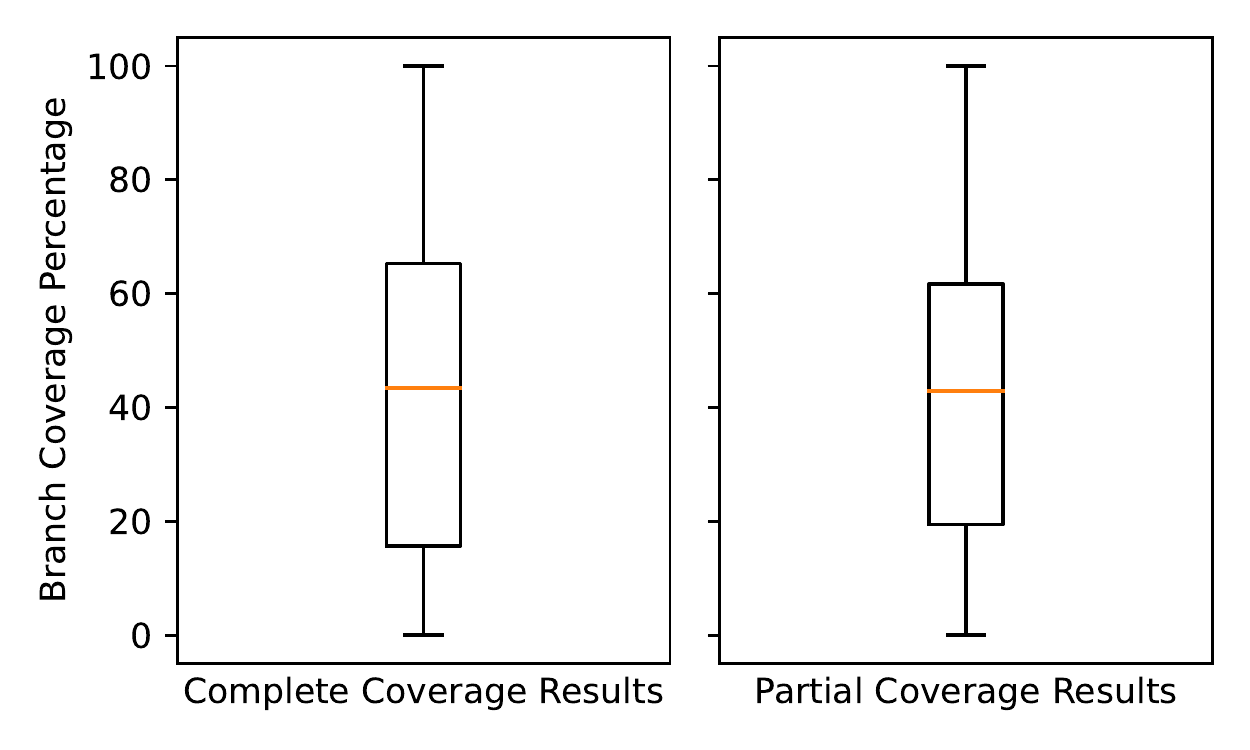}
    \caption{Branch coverage of projects with complete and partial coverage result.}
    \label{fig:branch-coverage}
\end{figure}

\noindent \textbf{Projects with partial coverage results.} For 138 projects we have obtained partial coverage results: 113 of them had at least one module with no test. In terms of why not all modules have JaCoCo output, we observe similar reasons as to our prior case of projects without any coverage result. If we compare the coverage levels of projects with partial and complete coverage results in Figure~\ref{fig:branch-coverage}, we observe a similar distribution. 

We also observe that from the total of 599 projects 389 (65\%) have at least one module with no tests (this is minus the projects without any tests since we removed them in the beginning). 

\medskip

Table~\ref{tab:8} provides an overview of how well projects are tested. Relying on the classification scheme from Heitlager \etal~\cite{DBLP:conf/quatic/HeitlagerKV07}, we observe that 2.9\% of projects fall in the category ``very high'' coverage, while a majority of projects are placed in the category ``low''. In fact, combing the categories ``low'' and ``very low'', we see that 68.4\% of projects fall into these two categories, giving an indication that not all projects are maximising their testing.

\begin{table}[!t]
    \centering
    \begin{smaller}
    \begin{tabular}{l|l|l}
    \toprule
        Reason              & \# & \%   \\ \midrule
        JaCoCo skipped/failed & 17 & 34\% \\
        Build failure         & 16 & 32\% \\
        Parsing error         & 3  & 6\%  \\
        Test failure          & 2  & 4\% \\
    \bottomrule
    \end{tabular}
    \end{smaller}
    \caption{\label{tab:jacoco-failure-reasons}Reasons of having no JaCoCo coverage result.}
    \end{table}

\begin{table}[!t]
\centering
\begin{smaller}
\begin{tabular}{c|c|c|c}
\toprule
    \textbf{Rank}  & \textbf{Code Coverage} & \textbf{\# of Projects} & \textbf{\% of Projects} \\ \midrule
    Very High (++) & 95-100\%               & 9                       & 2.9\%                   \\
    High (+)       & 80-95\%                & 31                      & 10\%                    \\
    Moderate (o)   & 60-80\%                & 58                      & 18.7\%                  \\
    Low ($-$)        & 20-60\%                & 129                     & 41.6\%                  \\
    Very Low ($--$)  & 0-20\%                 & 83                      & 26.8\%      \\            
\bottomrule
\end{tabular}
\end{smaller}
\caption{\label{tab:8}Ranking projects based on their code coverage using the classification scheme for unit test coverage proposed by Heitlager \etal~\cite{DBLP:conf/quatic/HeitlagerKV07}}
\end{table}

\begin{center}
    \fbox{\begin{minipage}{0.9\columnwidth}
        \textbf{Observation 6:} 68.4\% of projects among the ones we could calculate code coverage for, are doing very low or low testing.
        Also, only $\sim$13\% of them are doing very high or high testing. Moreover, 65\% of projects doing testing, do not do it for all modules of their project.
    \end{minipage}}
\end{center}

\revised{\subsubsection{Testing in Mature Projects}}
\revised{
We seek to understand whether and how following \textit{testing} as a quality 
assurance practice differs in mature projects compared to other projects 
in our dataset. 
Six projects out of the total 50 mature ones have code coverage results 
(6 out of 310 projects with JaCoCo results in Figure \ref{fig:testing-results-overview} are mature). 
Since this is a very low number to be representative of all mature projects, 
we manually looked all of 50 repositories to find their code coverage information. 
To this end, the first author checked repositories' \textit{README} files for a coverage 
badge\footnote{For more information on badges on GitHub, see: \url{https://github.com/dwyl/repo-badges}}, 
scanned through pull request comments, and looked for CI status checks on a project's GitHub page for
coverage information. Overall, we found code coverage information for 12 projects using this approach}
\footnote{\revised{Except one project using Jenkins service 
to measure branch coverage, other projects use Codecov service to measure their code coverage 
based on Codecov coverage metric: https://docs.codecov.com/docs/about-code-coverage, last visited September 7th, 2022.}}
\revised{If we combine the the information of these 12 projects with the JaCoCo results, 
we now have coverage information for 14 projects, i.e., for 4 projects we 
have both results of JaCoCo, and code coverage information from the GitHub page. We 
 picked the latter one for further analysis. Also, we checked to see if they have 
test suites in their repositories, and for all the mature projects we found test files 
indicating they have developed a test suite for their project.
}

\revised{
We observe a mean of 65.6\% code coverage among mature projects, with a median value of 63.2\%. 
This falls into the moderate level of testing, which is higher when we compare it to the mean of all other projects 
in our dataset, as they reach a low level of testing.
}

\revised{
\begin{center}
    \fbox{\begin{minipage}{0.9\columnwidth}
        \textbf{Observation 6*:} All mature projects have test suites in their repositories. Also, 
        on average they have 65.6\% code coverage (with median of 63.2\%) among the 14 projects
        that we could establish code coverage for. 
  	\end{minipage}}
\end{center}
}
\subsection{Quality Assurance Practices Put Together} \label{section:put-together}
In this section we set out to answer \textbf{RQ1.2}: \emph{Which quality assurance approaches are being used in conjunction?} In order to answer this question,
we use statistical tests to determine which quality assurance approaches that we previously discussed in isolation in Sections~\ref{section:localbuilds} through~\ref{section:results:testing} are used in conjunction in open source projects. 
More specifically, we conduct a statistical analysis to calculate the correlation between each of the practices using the results from previous sections: code coverage of 310 projects, code reviewing information of 1398 projects, ASAT usage and CI usage of all 1454 projects.
This analysis does raise the issue of data imbalance, as we have less projects for which we have code coverage data (310) than we have projects for which we have code reviewing information (1398). To address this issue, we have performed the statistical test on projects that are in the intersection for two quality assurance practices, i.e., those projects for which we have data of both. 

The correlation analysis enables us to better understand whether projects that use a quality assurance practice 
(intensively), are also more inclined to employ another quality assurance approach (intensively). Of importance 
to note here is that for two quality assurance practices, namely ASAT usage and CI usage, our data is categorical, 
i.e., a project \emph{is} or \emph{is not} using them. The other two practices, namely Testing and Code Reviews, 
are presented as continuous variables based on respectively the level of code coverage reached and the counts of 
comments/reviews that we observe. However, while for software testing we can make use of an established 
code coverage metric (branch coverage), for code reviewing we had to define a new metric to express the intensity 
of the code reviewing process (Equation~\ref{equ:code-review-rate} explained in Section~\ref{code-review-data}). 

In terms of the statistical tests, we use Cramer's V to measure correlation among categorical variables. Before applying Cramer's test, we first tested the independence of the variables using the Chi-square test. For the continuous variables vs. continuous variables and vs. categorical variables, we use respectively the Pearson and the Point Biserial correlation coefficient. The results of the correlation analysis are shown in Table~\ref{tab:correlation-analysis}.

\begin{figure}
    \centering
    \includegraphics[width=90mm]{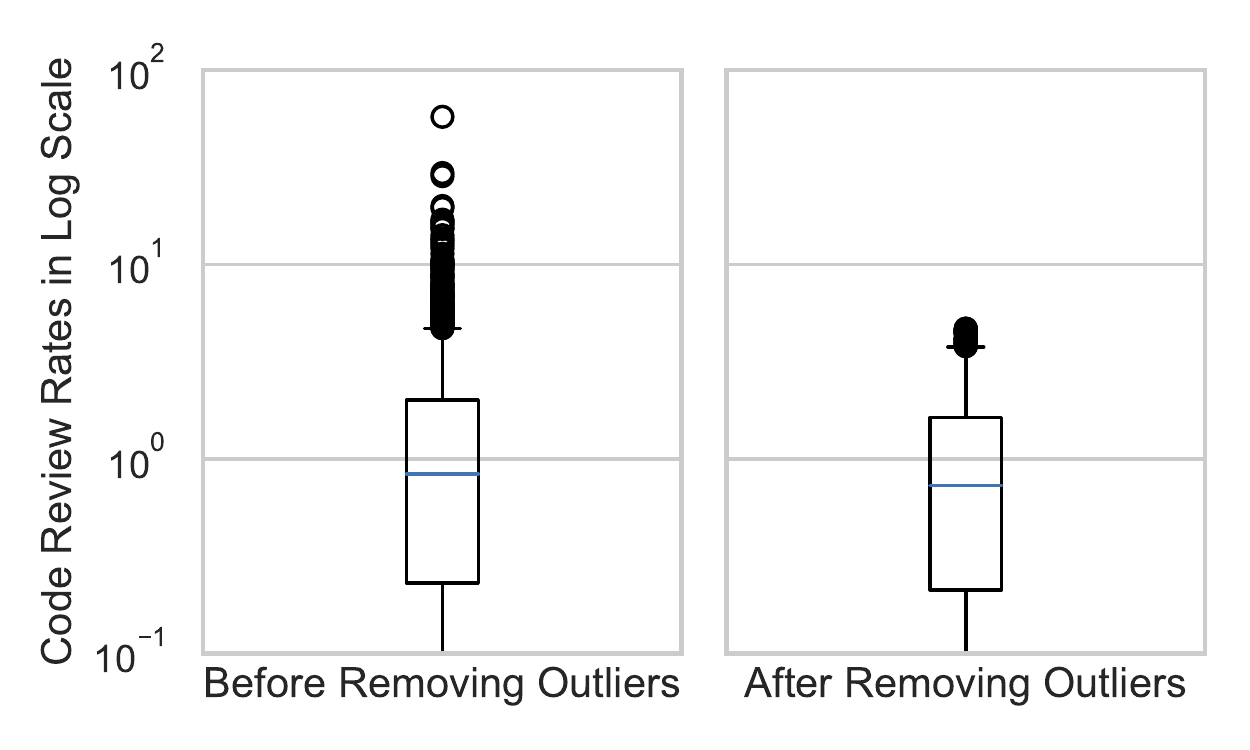}
    \caption{Code review rates before and after removing outliers.}
    \label{fig:cr-rate-before-after}
\end{figure}

\begin{figure}
    \centering
    \includegraphics[width=135mm]{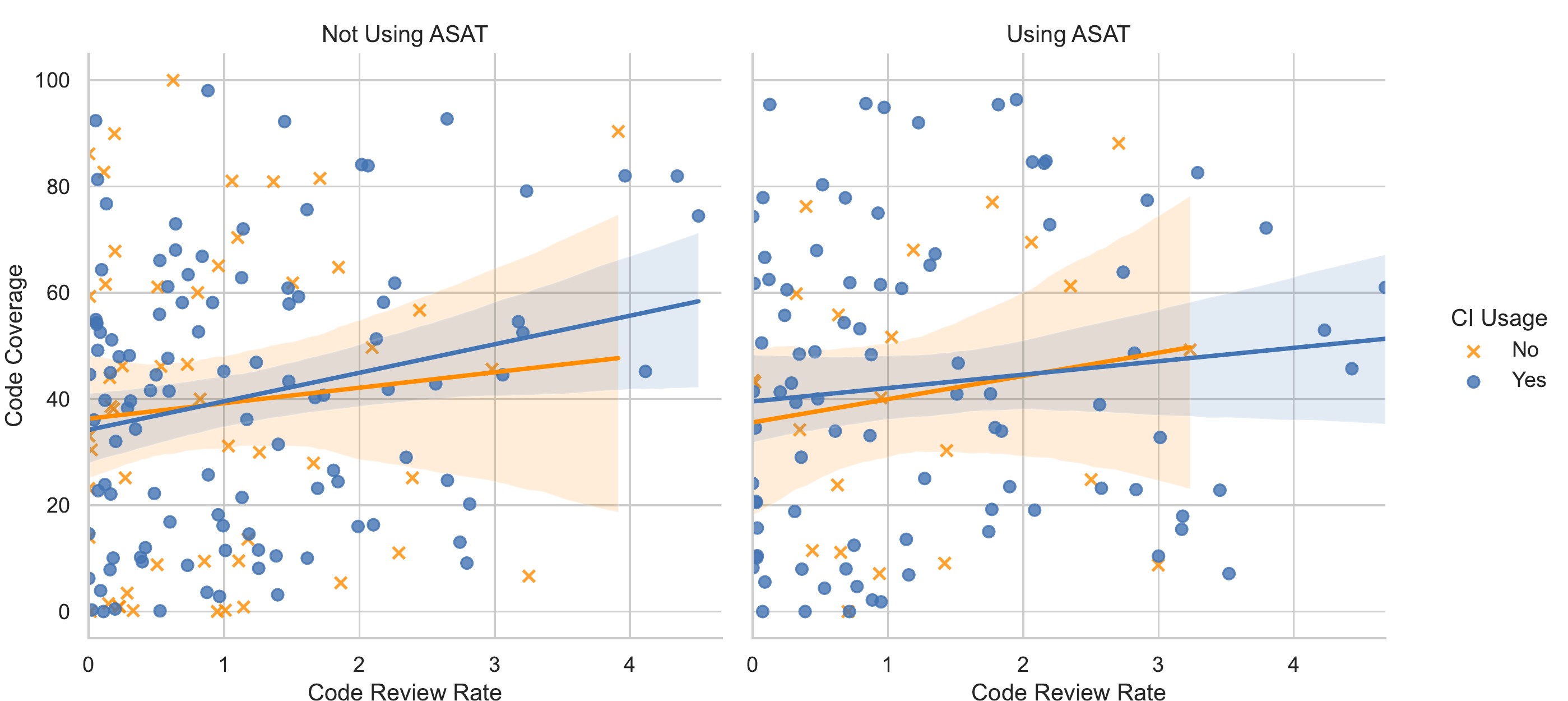}
    \caption{Relation between Testing and Code Reviewing with respect to usage of CI and ASAT. The outliers of code reviewing metric are excluded.}
    \label{fig:testing-cr-asat-ci}
\end{figure}

Table~\ref{tab:correlation-analysis} shows the results of the statistical tests. The results that are statistically 
significant are highlighted in light-grey. For our code review metric, since there were quite a lot of outliers 
(Figure~\ref{fig:cr-rate-before-after} shows the distribution of code review metric values before and after removing outliers), 
we used the IQR method to identify and then remove outliers~\cite{dekking2005modern}, which resulted in removing 106 projects 
(from a total of 1398 projects with code review rates). Therefore, there are two values for code review correlation analysis. 
Moreover, the same method was then used to remove the outliers of the code review metric in Figure~\ref{fig:testing-cr-asat-ci}. 
In contrast, the range of values for testing are fixed (0 to 100), and they are binary for ASAT/CI usage, so no outlier exists 
for other variables. 

\begin{table}[]
    \centering
    \renewcommand{\tabcolsep}{0.15cm}
    \begin{smaller}
    \begin{tabular}{c|llll}
    \toprule
    & \multicolumn{1}{c}{
        \begin{tabular}[c]{@{}c@{}}
            \textbf{Code}\\ 
            \textbf{Review}\end{tabular}} 
    & \multicolumn{1}{c}{
        \cellcolor{gray!35}\begin{tabular}[c]{@{}c@{}}
            \textbf{Code Review}\\ 
            \textbf{Without Outliers}\end{tabular}}
    & \multicolumn{1}{c}{\textbf{Testing}}
    & \multicolumn{1}{c}{\begin{tabular}[c]{@{}c@{}}\textbf{ASAT}\\ \textbf{Usage}\end{tabular}}
    \\ \midrule
    \textbf{Testing}
    & \cellcolor{gray!10}\begin{tabular}[c]{@{}l@{}}Corr = 0.29\\ p-value \textless 0.0001\end{tabular}
    & \cellcolor{gray!10}\begin{tabular}[c]{@{}l@{}}Corr = 0.15\\ p-value = 0.011\end{tabular}

    \\ \multicolumn{1}{c|}{\begin{tabular}[c]{@{}c@{}}\textbf{ASAT}\\ \textbf{Usage}\end{tabular}}
    & \begin{tabular}[c]{@{}l@{}}Corr = 0.019\\ p-value = 0.48 \end{tabular}
    & \cellcolor{gray!10}\begin{tabular}[c]{@{}l@{}}Corr = 0.12\\ p-value \textless 0.0001\end{tabular}
    & \begin{tabular}[c]{@{}l@{}}Corr = 0.0157\\ p-value = 0.79\end{tabular}

    \\ \multicolumn{1}{c|}{
        \begin{tabular}[c]{@{}c@{}}\textbf{CI}\\ \textbf{Usage}\end{tabular}}
    & \begin{tabular}[c]{@{}l@{}}Corr = 0.0317\\ p-value = 0.24 \end{tabular}
    & \cellcolor{gray!10}\begin{tabular}[c]{@{}l@{}}Corr = 0.10\\ p-value = 0.00024\end{tabular}
    & \begin{tabular}[c]{@{}l@{}}Corr = 0.0231\\ p-value = 0.69\end{tabular}
    & \cellcolor{gray!10}\begin{tabular}[c]{@{}l@{}}Cramer's V = 0.10 \\ p-value \textless 0.0001\end{tabular} \\
    \bottomrule
    \end{tabular}
\end{smaller}
\caption{\label{tab:correlation-analysis}Correlation analysis of quality assurance practices. Cells highlighted in light-grey are correlations that are significant.}
\end{table}

\medskip

\noindent \textbf{Code reviewing and testing} show a positive correlation with small effect size $r = 0.29$\footnote{For interpreting the strength of Pearson's correlation, we use 0 $\leq$ r $<$ 0.3 for \emph{small}, 0.3 $\leq$ r $<$ 0.5 for medium, and 0.5 $\leq$ r $<$ 1.0 for large effect size.} (also $r = 0.15$ when we remove outliers from code review rate). To better understand the small effect size, consider Figure~\ref{fig:testing-cr-asat-ci}. For both the projects that use ASATs and that do not use ASATs, we observe a concentration of projects that have a code review rate between 0 and 0.5 (the X-axis). Yet, that group of projects in that specific (relatively low) code review rate range show a widely dispersed code coverage level, going from 0\% all the way to 100\% branch coverage. This wide dispersal on the testing axis and clustering on the code reviewing axis (Y-axis) explains the small effect size. It also gives us the insight that projects that heavily invest in high levels of code coverage do not necessarily invest as strongly in code reviewing. 

\medskip

\noindent \textbf{Code reviewing and CI usage.} We observe a small positive correlation between code reviewing (without outliers) and CI usage $r = 0.10$. 
When we look at Figure~\ref{fig:testing-cr-asat-ci}, we need to consider both the left and right graphs and look at the X-axis for the code review rate and the spread of blue dots and orange crosses, signalling respectively the \emph{use} and \emph{non-use} of CI. For both the left and right graphs, we see an overall scattering of projects that use CI over the entire range of the code review rate. For the more dense code review rate interval [0,1.5], we visually observe a mix of projects that use or do not use CI. This scattering helps explain the small correlation that we obtained.

\medskip

\noindent \textbf{Code reviewing and ASAT usage.} We observe a small positive correlation between code reviewing (without outliers) and ASAT usage $r = 0.12$. 
When we look at Figure~\ref{fig:testing-cr-asat-ci}, we need to consider both the left (projects not using any ASATs) and right graphs (projects using ASATs) and look at the X-axis for the code review rate. We see an overall scattering of projects in the entire range of the code review rate. However, the density of projects with a code review rate between 0 and 2 is relatively higher in the left graph, On the other hand, we see more projects having a code review rate higher than 3 in the right graph. This density of projects in different ranges explains the small correlation that we have observed.

\medskip

\noindent \textbf{CI usage and ASAT usage} exhibit a weak, yet significant correlation $V = 0.10$\footnote{The effect size interpretation for Cramer's V is: 0 $\leq$ V $\leq$ 0.2 is weak, 0.2 $<$ V $\leq$ 0.6 is moderate, V $>$ 0.6 is strong}.  
When we turn our attention to Figure~\ref{fig:testing-cr-asat-ci}, we need to compare the distribution of blue and orange (respectively CI usage and no CI usage) dots/crosses in the left (not using an ASAT) and right (using an ASAT) graphs. Both left and right, we observe a multitude of projects that do and that do not use CI, with visually more orange crosses (for non CI usage) in the left most graph, signalling that no ASAT usage is more common amongst projects that do not use CI. To also see this numerically we divide the count of dots in the right graph by the total number of projects using an ASAT, and also divide the count of dots in the left graph by the total number of projects not using an ASAT. The results are respectively 0.79 and 0.67, confirming our earlier intuition that we have a higher percentage of projects that use CI among projects using ASATs, compared to projects not using any ASAT.

\medskip

Summarising the other relations from Table~\ref{tab:correlation-analysis}, we observe no other statistically significant correlations between the other quality assurance practices. 

\begin{center}
    \fbox{\begin{minipage}{0.9\columnwidth}
        \textbf{Observation 7:} Having high code coverage weakly correlates with doing more code reviewing in projects. We also observe weak correlations between (1) the usage of ASATs and CI usage, (2) the code review rate and CI usage, and (3) the code review rate and ASAT usage.
     \end{minipage}}
\end{center}

\section{Discussion} \label{section:discussion}
\subsection{Revisiting the Research Questions} \label{subsection:discussion:answering}
In this section, we revisit the research questions from Section~\ref{section:introduction}. 
In order to answer the over-arching research question

\begin{quote}
\textbf{RQ1} \emph{What is current state-of-the-practice in quality assurance in open source software development?}
\end{quote}
We first concentrate on providing a global overview of some of the most popular quality assurance practices in use, 
thereby answering \textbf{RQ1.1}

\begin{quote}
\textbf{RQ1.1} What is the prevalence of quality assurance approaches like software testing, modern code review, automated static analysis, and buildability?
\end{quote}

\revised{
As we mentioned before, RQ1.1 is aimed squarely at understanding the current state-of-the-practice with regard to 
the quality assurance practices of which we want to understand whether they are used in conjunction. We will now
discuss our findings for each of the quality assurance practices in isolation.}

\begin{description}
\item[Testing.]
From the original set of 1454 GitHub projects, we have been able to establish code coverage for 310 projects (21.3\%). Our results show that the median of their branch coverage is $\sim$43\%. Moreover, using Heitlager's classification scheme~\cite{DBLP:conf/quatic/HeitlagerKV07} of test coverage for projects, we establish that $\sim$68\% of them do \emph{very low}, or \emph{low} levels of testing. On the other hand, $\sim$13\% of projects reach \emph{high}, or \emph{very high} levels of code coverage. 

The reason why we could only establish code coverage for 21.3\% of the projects in our original dataset can be seen in Figure~\ref{fig:testing-results-overview} and can be summarized as a race of attrition because of projects' inherent attributes that make them ineligible for our study (28\%), issues of buildability (26.7\%), non-presence of tests and/or source code (4.1\%), and our inability to successfully establish code coverage results (19.8\%). 
\item[Code Review.]
For 1398 projects in our dataset we looked at the count of comments, and the count of reviews in their last 20 merged PRs. We have created a code review intensity metric, that expresses the amount of code reviewing normalised by lines of code changed in the PR. Subsequently, using Alves \etal's calibration method~\cite{DBLP:conf/icsm/AlvesYV10} we derived different categories of code reviewing intensity. We observe that projects having more than 50 reviews or 38 comments over their last 20 merged pull requests are among the projects doing more intense code reviewing. In contrast, projects having less than 38 reviews or 29 comments can be considered as projects doing relatively less code reviewing.
\item[ASAT Usage.] \label{disc-asats}
To study the usage of ASATs in GitHub projects, we make the assumption that the presence of an ASAT configuration file is an indication of actual ASAT usage. Under this assumption, we have observed that 38\% of the projects under observation use at least one ASAT. However, the presence of a configuration file does not causally imply the actual usage of these ASATs.
\item[Local Buildability.]
We set out to investigate whether the GitHub projects under study can be built locally out of the box. Our results indicate that $\sim$63\% of projects build successfully locally. 
\item[Continuous Integration.] 
We have used a new approach of studying usage of CI on GitHub, namely using GitHub's checks and statuses. This approach enables us to have a wide range of CI platforms in our study compared to previous studies that mainly focused on a single CI provider, e.g.,~\cite{DBLP:conf/msr/BellerGZ17}. We find evidence that $\sim$63\% of overall 1454 projects use CI with $\sim$76\% of them having a successful state. This also indicates that for 24\% of the projects, the final commit that is part of our study led to a failing CI build. Moreover, we observe different strategies in making use of GitHub features (GitHub Actions, and Statuses) to implement CI status checks among projects (e.g., having a few vs. many CI status checks with many different CI platforms or services). In future work, we intend to study the reasons for using multiple CI status checks and platforms. 

\end{description}

Now that we have an understanding of how these quality assurance practices 
are used in isolation, we turn our attention to:
\begin{quote}
\textbf{RQ1.2} Which quality assurance approaches are being used in conjunction?
\end{quote}

\noindent \textbf{Code Review and Testing.}
We found statistically significant evidence that higher levels of branch coverage correlate with higher levels of the code review intensity metric with small effect size. 
This indicates that projects that show a higher level of branch coverage tend to have more code reviewing performed.

\medskip

\noindent \textbf{Code Review and ASAT/CI usage.}
Using our defined categories of code reviewing intensity, if we look at projects with a very high level of code reviewing, e.g., more than 73 reviews or 56 comments in their last 20 merged PRs (239 projects), for 49\% of them we find evidence of ASAT usage, and for 65\% we see evidence of CI usage. On the other hand if we look at projects with a low level of code reviewing, e.g., less than 38 reviews and less than 29 reviews in their last 20 merged PRs (781 projects), for 32\% of them we find evidence of ASAT usage, while for 61\% we see evidence of CI usage.

This observation shows that projects with a higher code review intensity tend to use ASAT and CI more when comparing to the ones with lower code review intensity. We also saw statistically significant correlation between our code reviewing intensity metric and ASAT/CI usage when outliers are removed, which strengthens this observation. 

\medskip

\noindent \textbf{ASAT usage and CI Usage.}
Finding evidence of ASAT usage in projects does not imply that these ASATs are actually used and integrated in the CI workflow. However, we did observe a statistically significant, yet weak correlation between their usage, which potentially indicate that projects use both at the same time.

\medskip

\begin{quote}
\textbf{\revised{RQ1.3:}} \revised{How do mature projects follow quality assurance practices? }
\end{quote}

\revised{To summarize our findings with regard to the quality assurance practices in more mature projects, compared to our overall dataset, we can state that:}
\begin{enumerate*}
\item \revised{\textbf{local buildability} is almost the same, which also indicates that keeping the latest version buildable is also a challenge for mature Java projects,}
\item \revised{\textbf{ASAT usage} is 22\% higher in the more mature projects,}
\item \revised{the \textbf{CI usage} ratio is similar over both sets, but CI states are less successful in the more mature projects, }
\item \revised{the \textbf{code review} intensity is either high or very high for around half of the more mature projects, while over the entire dataset the projects with 
the 20\% highest code review intensity are in these categories (see Table~\ref{tab:7} for thresholds). This is an indication that more mature projects typically do code reviews at a higher intensity, }
\item \revised{finally, test suites of more mature projects seemingly have higher \textbf{code coverage}. Moreover, based on the existence of unit tests in \emph{all} mature project repositories, we assume they all do testing.}
\end{enumerate*}

\subsection{\revised{Difficulties encountered during our investigation}}
\revised{We now give an overview of the challenges and obstacles that we have encountered during our study. We hypothesize that 
understanding and potentially alleviating these challenges in the future can be useful for researchers, and perhaps also practitioners. 
}

\noindent \textbf{Difficulty of creating an environment for local build success.}
We could not build $\sim47\%$ of projects out of the box. Compilation, dependency, and configuration 
errors are the top three reasons of the build failures. Also, other studies have reported that building 
Java projects is not always straightforward. Hasan \etal~could not build 46\% of the projects in their 
study~\cite{DBLP:conf/esem/HassanMLW17}. Sulir \etal~have respectively observed a 38\% and 59\% build 
failure percentage in their local build results dataset in their 2017 and 2020 
studies~\cite{DBLP:journals/corr/abs-1712-01024, DBLP:journals/data/SulirBMCJ20}.

This number of build failures shows that it is non-trivial to create an easy-to-build environment for Java projects. 
This has far-reaching consequences for both \textbf{practitioners} and \textbf{researchers}. For \textbf{practitioners}, 
this means that developers that want to contribute to an open source project might find it difficult to build the project 
on their development machine. On the other hand, \textbf{researchers} might be hindered to conduct their studies on a 
large variety of projects, or on multiple versions of a project. Several studies have investigated the causes of 
build failures~\cite{DBLP:conf/msr/RauschHL017, DBLP:conf/sigsoft/LouCCHZ20}, while other studies have tried to 
devise approaches to tackle the underlying 
issues~\cite{DBLP:conf/kbse/Hassan19, DBLP:conf/wcre/MachoM018, DBLP:conf/sigsoft/LouCCHZ20, DBLP:journals/jss/BarrakEAK21}. 
In an ideal world, creating an environment (e.g., a docker container) for building a large number of projects would benefit 
\textbf{researchers} to conduct studies that need to build projects as a prerequisite (e.g., dynamic analysis, measuring code coverage, 
etc.).

Interestingly, our results indicate that if project maintainers want more confidence about their projects' 
buildability on local machines, they should consider doing continuous integration, because CI increases the 
likelihood of a successful local build (Cramer's V $\simeq$ 0.08). And also consider having a successful CI build 
for the latest release of the project. 
Because, CI builds match with local build outcomes in 64\% of projects, and there is a statistically significant 
correlation between them (Cramer's V $\simeq$ 0.10).

\medskip

\noindent \textbf{Measuring code coverage.}
Measuring code coverage of many projects automatically is a challenge for \textbf{researchers}. 
One prerequisite of measuring code coverage of a Java project is building it, 
and we already discussed that creating an environment for building a large number of 
projects successfully is a challenge. The next step in the measurement of code coverage 
is to use a code coverage measurement tool. Our observations are that, first, most projects 
do not have their code coverage details available in their repository. So, in most cases researchers 
need to measure projects' code coverage themselves.
Second, automatically generating code coverage information for a large number of projects 
is a cumbersome task. This is because of the difficulty of integrating
a code coverage tool like JaCoCo to generate code coverage results, and also successfully building 
the project with running all tests. 
Particularly, this is why we were unable to collect coverage measurements for $\sim$48\% of projects 
when trying to measure their code coverage. Third, many projects have several modules (subprojects), 
and they do not have unit tests for all of them. This issue makes it difficult to have an idea of overall 
code coverage since some parts of the software project miss tests. This third issue in particular, is 
also an issue for \textbf{practitioners}, as it leaves them blind with regard to the level of test coverage of 
specific submodules.

\medskip

\noindent \textbf{Measuring the quality of code review.}
In this paper we suggested a code review metric to have an idea of the intensity of reviews taking place during 
pull requests' discussions. Then we looked at the correlation between testing, ASAT usage, and CI usage with 
code review intensity. However, the intensity of code review is different from the quality of the code review. 
Despite the fact that code review quality has been studied in many studies, there is no obvious way to 
quantitatively measure it~\cite{DBLP:conf/icse/KononenkoBG16, DBLP:journals/jss/BarrakEAK21, DBLP:conf/icsm/KononenkoBGCG15}. 
Having a measurement of the quality of the code review would help \textbf{researchers} when comparing projects' 
code review and when studying its correlation with other quality assurance practices. Also, with a 
code quality measurement \textbf{practitioners} have a concrete idea of the state of the quality of the code reviews 
taking place in their project, and it enables them to improve their code review practices. 

\medskip

\noindent \textbf{ASATs actual usage}
As we pointed out before (see Section~\ref{disc-asats}), the presence of a configuration file does not causally imply the \emph{actual usage} of these ASATs. 
Vassallo \etal~found that developers use ASATs in different contexts: local programming, code review, and CI~\cite{DBLP:conf/wcre/VassalloPPPZG18}. 
So, to have a notion of ASATs actual usage we need to look at them in all three different contexts, 
which is a challenge for researchers, keeping in mind that different projects use different CI platforms. 
We have looked at how other studies approach the study of actual usage of ASATs: 
Beller \etal~investigated the use of ASATs in 122 projects by manual analysis of configuration files and then surveying the contributors of those 
projects, asking about the actual usage of ASATs \cite{DBLP:conf/wcre/BellerBMZ16}. In another work Zampetti \etal~studied 
the actual use of ASCATs (Automated Static Code Analysis Tools) in the CI environment~\cite{DBLP:conf/msr/ZampettiSOCP17}. 
In their study, they analyzed Travis CI builds and jobs of 20 projects.

For \textbf{researchers}, observing and understanding the actual usage of ASATs at scale remains a challenge. 
A key challenge is that ASATs can be used in multiple contexts (per the observation of Vassallo \etal~\cite{DBLP:conf/wcre/VassalloPPPZG18}), 
thus requiring multiple observation points. Another challenge is that in order to obtain meaningful insights, 
the dataset would need to be sufficiently large, implying that we need an automated observation tool to perform 
a longitudinal analysis on multiple projects.

\medskip

\noindent \textbf{GitHub Actions best practices}
To study the use of CI, we put our focus on the status checks of a project's latest commit. 
We found that projects have different approach when configuring their CI using GitHub features: 
GitHub Actions (checks run using GitHub's apps) \& Commit Statuses (statuses marked by external services). 
There are projects with more than 100 status checks, and many others with only 1 check / status. Also, in 
some cases, many GitHub apps are used in parallel to run checks. 
Here we highlight the challenge for \textbf{practitioners}: configuring their CI using checks (GHA) and statuses. 
Including choosing the right (combination of) GitHub app(s) for these checks. 

Kinsman \etal~have shown that developers have a positive perception of GHA~\cite{DBLP:conf/msr/KinsmanWGT21}, 
and in a related work about CI services on GitHub, Golzadeh \etal~found that GHA has become the dominant CI 
service 18 months after its introduction~\cite{golzadehrise}, which strengthens the need to address 
the mentioned challenges by offering best practices to use GHA. 

In another work, Chen \etal~took a different path to study the usage of GHA focusing on the projects' 
GHA YAML file (under the workflow path of projects adopting GHA). They similarly found a median of 3 for 
steps (median of 3 for status checks in our study) used in jobs (with average of 4.7 vs. average of 7.35 
status checks in our study), and highlighted the challenge for less experienced developers to configure a 
GHA workflow when facing with workflows, jobs, and steps with many alternatives.

\subsection{The Bigger Context of Quality Assurance Practices}
In Section~\ref{subsection:discussion:answering} we have revisited research questions \textbf{RQ1.1}, \textbf{RQ1.2} and \revised{\textbf{RQ1.3}} and gave concrete answer to them. However, we have not answered the over-arching RQ1 yet. In order to do so, we need to look at the bigger context. 

\begin{quote}
    \textbf{RQ1} \emph{What is current state-of-the-practice in quality assurance in open source software development?}
\end{quote}

First of all, we have observed a very diverse level of adoption, or intensity of adoption, of the individual quality assurance practices among our set of 1454 open source software projects on GitHub. We have observed projects that do not invest in any of the practices, but we have also seen projects investing in all, or at least most practices with relatively high intensity. Also, when we look at mature projects we see that they generally tend to invest more in their quality assurance practices. 

When it comes to using quality assurance practices in conjunction, we have observed weak correlation among some quality assurance practices in projects. But generally speaking, projects do not follow all quality assurance practices together with high intensity. This raises interesting questions for follow-up research that is more qualitative in nature, because some of the studied quality assurance practices are said to be complementary, using them in conjunction would strengthen the overall quality perspective, and might also reduce quality assurance efforts (e.g., an ASAT might pick up code review comments, which would ease the burden of the reviewer~\cite{DBLP:conf/wcre/PanichellaAPA15}).

\medskip

\revised{\subsection{Related Work}}
\revised{In this section we compare the results of our study with other related studies.
}

\revised{\subsubsection{Quality Assurance Practices in Isolation}}
\revised{Most of the related work has studied each of the quality assurance practices in isolation. 
Here we look how our results for the usage 
of each of these quality assurance practices in isolation compare to other related studies.}

\revised{\noindent \textbf{Buildability.}
Hassan \etal~could build 54\% of the projects in their dataset using the default build commands of build tools. This 
ratio is 9\% lower than our 63\% of projects that build successfully out of the box~\cite{DBLP:conf/esem/HassanMLW17}. 
While their methodology for establishing buildability is very similar to ours, their dataset comprises fewer projects (200 vs. 1456). Of importance to note is that their dataset also comprises projects that us the Ant build system. Moreover, they 
have discussed that 52 of the build failures can be resolved automatically. Following this work, Hassan and Wang 
created HireBuild as an automatic approach to fix reproducible build failures (which is a history-driven approach, thus could not be used 
in our study to address build failure issues)~\cite{DBLP:conf/esem/HassanW17}. Both of the aforementioned studies highlight 
the challenge of facing build failures. 
Sulir \etal~have respectively observed 
a 38\% and 59\% build failure percentage in their local build results dataset in their 2017 and 2020 
studies~\cite{DBLP:journals/corr/abs-1712-01024, DBLP:journals/data/SulirBMCJ20}. Similar to our observations, both of their studies report \textit{`dependencies'} and \textit{`Java compilation'} as two of the top categories of 
errors leading to build failure of the projects.}

\revised{\noindent \textbf{Testing.}
Our goal of measuring code coverage of projects was to see how testing is being done in conjunction with 
other quality assurance practices. To the best of our knowledge no other studies have done such large-scale study of code coverage 
of open-source projects' unit tests for this particular goal. Other studies have looked at code coverage of projects to study its evolution~\cite{DBLP:conf/icse/GopinathJG14}, or 
its relation to software quality~\cite{DBLP:journals/tr/KochharLLN17}. Hilton \etal~have taken a similar approach collecting 
code coverage information for 29 projects using JaCoCo~\cite{DBLP:conf/kbse/Hilton0M18}. However, their study was aimed at studying the coverage evolution, and thus they focused on obtaining 250 successful builds per project for a limited set of projects, instead of single successful builds of many projects in our study. 
Overall, for the 29 Java projects that they have considered, six of them had $<$60\% coverage and only one had $<$20\%. The average  
code coverage percentage for all projects in their study was 76\%
~\cite{DBLP:conf/icse/GopinathJG14}. 
Kochhar \etal~have used Sonar to measure code coverage of 100 open-source Java projects. 8\% of the projects in their dataset had higher 
than 75\% coverage. Also, 69\% of them had a coverage of less than 50\%. The results of the coverage of their projects are similar to 
ours, however they used a different coverage metric which is a combination of branch and statement coverage. 
One of the main reasons of the difference in results of our study comparing 
to others is that we have used a relatively larger set of projects. Addressing the difficulty of obtaining code coverage measures for researchers, would open up the way to more studies of the state of the practice in testing.}

\revised{\noindent \textbf{Continuous Integration.}
One of the main differences between our CI usage study and other related works lies in our approach to detect CI usage. 
Many other studies have used either the TravisTorrent dataset~\cite{DBLP:conf/msr/BellerGZ17a}, or the Travis-CI API for this matter
~\cite{DBLP:conf/wcre/CasseeVS20, DBLP:conf/msr/BernardoCK18, DBLP:conf/sigsoft/RahmanAKS18, DBLP:conf/sigsoft/VasilescuYWDF15, DBLP:conf/kbse/ZhaoSZFV17, DBLP:conf/icsm/NeryCK19, DBLP:conf/kbse/HiltonTHMD16}. 
Other studies have looked at Jenkins~\cite{DBLP:conf/sigsoft/RahmanAKS18, DBLP:conf/sigsoft/VasilescuYWDF15}, 
CircleCI~\cite{gallaba2022lessons, DBLP:conf/sigsoft/RahmanAKS18, DBLP:conf/kbse/HiltonTHMD16}, 
SonarCloud~\cite{gallaba2022lessons}, AppVeyor~\cite{DBLP:conf/kbse/HiltonTHMD16}, or Werker~\cite{DBLP:conf/kbse/HiltonTHMD16} as the CI services 
configured with GitHub projects in their studies. 
However, the introduction of GHA aligns with a decreasing growth rate 
of CI usage for Travis and CircleCI, two of the most popular CI platforms before the introduction of GHA~\cite{golzadehrise}. 
Considering this fact, we have used a new approach to detect CI usage in GitHub projects 
regardless of the CI platform that they use. This approach gave us the ability to study more projects.  
}

\revised{Hilton \etal~have observed a 40\% CI usage in their corpus of $\sim$34K projects. When they only considered the most popular projects in their corpus, they observe a 70\% usage ratio
~\cite{DBLP:conf/kbse/HiltonTHMD16}. 
In order to make a fair comparison, we only considered the Java projects in Hilton \etal's dataset when comparing to our results: they
had a 35\% CI usage among their 3371 Java projects, while we have observed a 63\% usage rate some five years later. 
This point also underlines the need for more research on GHA adoption among projects as the most popular CI platform among open-source projects to 
highlight best practices as mentioned in Section~\ref{subsection:discussion:answering}.}

\revised{\noindent \textbf{Code Review.}
McIntosh \etal~have found that reviews without discussion are associated with higher post-release defect counts~\cite{DBLP:conf/msr/McIntoshKAH14}. Accordingly, we use count of 
comments and reviews in our study as a proxy for code review intensity. They have also studied the effect of code review coverage metrics on 
software quality, however this contributes less than code review participation metrics to software quality defect models. 
Bosu \etal~have identified a set of factors affecting the usefulness of code reviews~\cite{DBLP:conf/msr/BosuGB15}. They recommend submitting smaller and incremental 
changes. Therefore, using this insight, we normalized count of comments and reviews in code review discussions by the number of lines of 
code changed in each pull request. 
To the best of our knowledge no other study has suggested any quantitative code review intensity / quality metric for projects on GitHub. 
}

\revised{\noindent \textbf{ASAT Usage.}
Beller \etal~did an analysis of the state of ASAT usage in open source software. They have found that 60\% of the popular OSS projects 
have automated static analysis configured, however they usually only use one ASAT and do not integrate it into their workflows~\cite{DBLP:conf/wcre/BellerBMZ16}. 
Comparing to their results, our study shows 38\% of ASAT usage among projects which increases to 70\% by considering only the mature ones. 
Checking whether ASATs are integrated into the workflows of projects is out of the scope of our study. But since we also see an increase in CI usage 
among mature projects, integrating ASATs in CI workflows can be a factor in this observation.
Other related studies have looked at the reasons why developers use ASATs~\cite{DBLP:journals/tse/DoWA22}, how they engage with them in different 
contexts~\cite{DBLP:journals/ese/VassalloPPPGZ20}, and how developers employ them in 
their CI pipelines~\cite{DBLP:conf/msr/ZampettiSOCP17}. We will discuss more about their results in the next part when discussing related work of different quality assurance practices being used in conjunction.}

\revised{\subsubsection{Quality Assurance Practices in Conjunction}}
\revised{To the best of our knowledge our work is the first study into how a set of quality assurance practices are being used in conjunction. Other related work has looked at the influence of one practice 
on another by focusing on studying the relation between them. Here we look at the results of these studies and compare them 
to ours.}

\revised{Literature on ASATs suggests to configure ASATs in development workflows, e.g., CI pipelines~\cite{DBLP:conf/wcre/BellerBMZ16}. In particular, Zampetti \etal~have studied how 20 popular Java open source projects use ASATs in their CI. Specifically, they have investigated which ASAT tools are being used and how they are configured for CI, and what types of issues make the build fail or raise warnings. They have observed that CI builds typically break due to ASATs detecting non-adherence to coding standards~\cite{DBLP:conf/msr/ZampettiSOCP17}. 
In our analysis, we have observed a statistically significant, yet weak correlation between CI usage and ASAT usage, 
indicating that ASATs are indeed potentially configured in CI pipelines. Also, the ratios of CI and ASAT usage both increase when we filter for mature projects. 
}

\revised{In another related work, Zampetti \etal~studied the interplay between pull request reviews and CI builds. They found 
that pull request discussions mainly focus on testing, static analysis problems, and also about CI configurations~\cite{DBLP:conf/wcre/ZampettiBCP19}. 
This seemingly confirms that some quality assurance approaches are used in conjunction, in particular code reviewing and CI builds.}

\revised{Cassee \etal~found CI to be a silent helper of code reviews by saving time. They specifically found that the number of comments in code reviews 
decreases after adopting CI, without any change in the number of changes per code review~\cite{DBLP:conf/wcre/CasseeVS20}. Our study has indicated a statistically significant, yet 
weak correlation between CI usage and intensity of code reviews, which means that projects with high code review intensity tend to use CI more when comparing 
to the ones with lower code review intensity. 
So from Cassee \etal~we learn that using CI can be a time saver, while our observations point to projects with high code review intensity 
already to be using CI. }

\revised{In another similar study, Rahman \etal~found that automated builds very likely affect code review participation. They stated that 
automated builds along with their outcomes might trigger further code reviews~\cite{DBLP:conf/msr/RahmanR17}. This can also be a reason why we see a statistically significant 
correlation between code review intensity and CI usage. We encourage future research to qualitatively study the relation between CI usage and 
code review intensity.}

\revised{Wang \etal~studied the combined use of testing and CI in open source projects by focusing on how test automation maturity 
can improve product quality. They found product quality improvement and release cycle acceleration as the potential benefits of 
test automation maturity~\cite{DBLP:journals/jss/WangMLM22}. In support of our own investigation, the study by Wang \etal~confirms
the importance of using specific quality assurance practices in conjunction. 
}

\revised{\subsection{Implications and Future Work}}
\revised{
We envision that within the software engineering community there are two groups that can benefit from the results of 
our study, namely researchers and educators. We acknowledge that we need additional insights from future research to bring value to practitioners as well.}

\smallskip
\noindent\revised{\textbf{Researchers.} Our investigation has shown that among most quality assurance practices no or only a weak correlation can be observed. 
This calls for a deeper understanding of the complementarity of the approaches. Simultaneously, we need to better comprehend why practitioners choose between 
practices, and not do multiple of them at a similar level of intensity. We hypothesize that practitioners need a better grasp of how effective each of the quality assurance 
practices is to reach a certain level of quality. In this area, we also refer to the work of Cassee \etal~which shows that CI can reduce the effort required from code reviews~\cite{DBLP:conf/wcre/CasseeVS20}. We make an explicit call to arms, to further investigate the effectiveness and efficiency of using quality assurance practices in conjunction.}

\revised{Another important and timely step for researchers to make is to establish and measure the quality and intensity of code reviewing. Additionally, with the 
growth in popularity of GitHub Actions, a deeper investigation into how this platform is used and should be used is also an urgent concern for future research.
Furthermore, we propose to replicate our study for other programming languages, e.g., Python, Javascript, to get a better understanding of the state-of-practice 
in other communities. 
We also envision that it can be interesting to investigate how practitioners' behaviour in applying quality assurance practices changes, when they are better
informed in terms of the effectiveness of the quality assurance approaches, both when used in isolation and in conjunction. 
Finally, we consider that another interesting path for future research is to investigate the challenges that open-source developers encounter when trying to
contribute to open-source software projects.
}

\smallskip
\noindent\revised{\textbf{Educators.} Based on our observations, we would like to emphasize the importance of teaching about all of these quality assurance practices, thereby also 
highlighting their complementarity so that these future practitioners can make well thought through trade-offs in terms of quality assurance. Of importance to underline towards students
is our observation that mature projects tend to combine quality assurance practices more intensely. Another important element to highlight here is
that students should be made aware of the importance of leaving the project in a buildable state, as to enable newcomers to easily contribute to the project. }

\subsection{Threats to Validity}

\noindent
\textbf{External validity.}
We have used the GHS dataset~\cite{Dabic:msr2021data} to select 1454 projects from GitHub that are written in the Java programming language, had a sufficiently large number of pull requests (200), minimally 10 contributors, and are popular ($>$100 stars). For some of our investigations, we could not use the entire dataset (e.g., because there was no test code, or we could not measure test coverage). While we have taken great care not to introduce any bias in the projects that we study, the filtering conditions that we use might impact the external validity of our results. We urge the research community to replicate our studies, e.g., with more software projects, also in the context of different programming languages.

Another threat is that we only consider 3 different ASATs. While these are the more popular ASATs for the Java programming language~\cite{DBLP:conf/wcre/BellerBMZ16}, we cannot claim that our findings generalize to other ASATs or ASATs in other programming languages.

While we have looked at multiple CI platforms, and went beyond the TravisTorrent dataset in terms of scope~\cite{DBLP:conf/msr/BellerGZ17a}, we have only considered CI tools that are configured in GitHub. We cannot generalize our findings to CI systems that are used outside of GitHub.

\medskip
\noindent
\textbf{Construct validity.}
In Section~\ref{section:results:testing} we study test quality through the \emph{branch test coverage} metric. While test coverage metrics provide a good image of which parts of a code base are not covered by tests, it does not provide insight into how good the tests actually are at detecting faults. Running a full mutation analysis would give us more insight here, but is also a computationally very costly operation. 

In Section~\ref{code-review-data} we create a code review intensity metric, that normalises the number of code review comments per line of code. While we have validated this metric, we do work under the assumption that more code review comments per line of code are better, but we should also take into consideration that a team of experienced contributors might produce higher quality code in the first place, thus requiring less suggestions on how to improve. Future research should look into the relationship between developer experience and code review intensity.

\revised{In order to determine local buildability, we relied on a ``minimal environment'' for building the projects out of the box (without any special configuration, reading project specific documents, and environment setup). More specifically, we used a Docker container of Ubuntu 21.04 with the Java Development Kit (Open JDK 11) and build tools such as Maven and Gradle. We do realise that some projects have special needs in order to be built, but in our aim of investigating a large number of projects automatically, this was a conscience trade-off.}

\section{Conclusion} \label{section:conclusion} 

In this study, we set out to understand the state-of-the-practice in quality assurance in open source software development. In particular, we have focused on software testing, modern code review, automated static analysis, 
and build automation. We acknowledge that each of these practices have been studied in-depth in isolation, and as such, the major contribution of this study is that we shed light on how the 
quality assurance practices are used in conjunction. We have collected quality assurance related information from 1,454 open source Java projects from GitHub. This information includes their 
local buildability outcomes, code coverage results, discussion \& review details at the pull request level, usage of automated static analysis tools, and continuous integration related status checks. 

In our main research question \textbf{RQ1} \emph{(What is current state-of-the-practice in quality assurance in open source software development?)} we first looked into each of the quality assurance practices in isolation, observing a wide variety of their intensity of usage. Moving the goalpost to understand how these quality assurance practices are used together, we observe 
weak correlation in our set of projects when it comes to (1) code coverage and code review intensity, (2) usage of ASATs and usage of CI, (3) code review intensity and CI usage, and (4) code review intensity and ASAT usage. 
\revised{We also specifically zoomed in on the more mature projects in our dataset, and in general we observe that these are more intense in their application of the aforementioned quality assurance practices, with more focus on their ASAT usage, and code reviewing, but no strong change in their CI usage.}

\revised{In our investigation of the combined usage of quality assurance practices, we have overall observed weak correlations. This calls for a future investigation that deeply tries to understand the complementarity of the quality assurance approaches that we consider, and the trade-offs that developers face when choosing between them.}

\section*{Acknowledgements}
This research was partially funded by the Dutch science foundation NWO through the Vici ``TestShift'' grant (No. VI.C.182.032).

\balance
\bibliographystyle{elsarticle-num}
\bibliography{references}

\begin{thebibliography}{10}
\expandafter\ifx\csname url\endcsname\relax
  \def\url#1{\texttt{#1}}\fi
\expandafter\ifx\csname urlprefix\endcsname\relax\def\urlprefix{URL }\fi
\expandafter\ifx\csname href\endcsname\relax
  \def\href#1#2{#2} \def\path#1{#1}\fi

\bibitem{TechCrunch2016}
J.~Patel, Software is still eating the world,
  \url{https://techcrunch.com/2016/06/07/software-is-eating-the-world-5-years-later/?guccounter=1&guce_referrer=aHR0cHM6Ly93d3cuZ29vZ2xlLmNvbS8&guce_referrer_sig=AQAAADIqx8LBuU1uKI03errh0RlYZjGsX_ZK76KVXqy3KqGkv3xyyXVrxi-46rFMEmZaBV4Na7Cm2lYLUC_QcKfhx0-njTwVR8XKsjkrDvNC9CoaHj4L9SLucX6hkJUcxl-rhBjsxcATrgy0yFSpYqhGgq9yKJY5F8mVs9sSV-AADaO-},
  last visited May 20th, 2022 (2016).

\bibitem{JazayeriASE2004}
M.~Jazayeri, The education of a software engineer, in: Proc. International
  Conference on Automated Software Engineering (ASE), IEEE, USA, 2004.

\bibitem{DBLP:conf/icse/KoDD14}
A.~J. Ko, B.~Dosono, N.~Duriseti, Thirty years of software problems in the
  news, in: Proceedings of the 7th International Workshop on Cooperative and
  Human Aspects of Software Engineering ({CHASE}), {ACM}, 2014, pp. 32--39.

\bibitem{Aniche2022}
M.~Aniche, Effective Software Testing: A Developer's Guide, Manning
  Publications, 2022.

\bibitem{DBLP:journals/tse/KameiSAHMSU13}
Y.~Kamei, E.~Shihab, B.~Adams, A.~E. Hassan, A.~Mockus, A.~Sinha, N.~Ubayashi,
  A large-scale empirical study of just-in-time quality assurance, {IEEE}
  Trans. Software Eng. 39~(6) (2013) 757--773.

\bibitem{DBLP:conf/icse/BacchelliB13}
A.~Bacchelli, C.~Bird, Expectations, outcomes, and challenges of modern code
  review, in: 35th International Conference on Software Engineering ({ICSE}),
  {IEEE}, 2013, pp. 712--721.

\bibitem{DBLP:conf/wcre/BellerBMZ16}
M.~Beller, R.~Bholanath, S.~McIntosh, A.~Zaidman, Analyzing the state of static
  analysis: {A} large-scale evaluation in open source software, in: {IEEE} 23rd
  International Conference on Software Analysis, Evolution, and Reengineering
  ({SANER}), {IEEE}, 2016, pp. 470--481.

\bibitem{DBLP:journals/ese/VassalloPPPGZ20}
C.~Vassallo, S.~Panichella, F.~Palomba, S.~Proksch, H.~C. Gall, A.~Zaidman, How
  developers engage with static analysis tools in different contexts, Empir.
  Softw. Eng. 25~(2) (2020) 1419--1457.

\bibitem{DBLP:conf/msr/BellerGZ17}
M.~Beller, G.~Gousios, A.~Zaidman, Oops, my tests broke the build: an
  explorative analysis of {Travis} {CI} with {GitHub}, in: Proceedings of the
  International Conference on Mining Software Repositories ({MSR}), {IEEE},
  2017, pp. 356--367.

\bibitem{RahmanRCOSE2017}
A.~Rahman, A.~Partho, D.~Meder, L.~Williams, Which factors influence
  practitioners' usage of build automation tools?, in: International Workshop
  on Rapid Continuous Software Engineering (RCoSE), 2017, pp. 20--26.

\bibitem{DBLP:conf/msr/RauschHL017}
T.~Rausch, W.~Hummer, P.~Leitner, S.~Schulte, An empirical analysis of build
  failures in the continuous integration workflows of java-based open-source
  software, in: Proceedings International Conference on Mining Software
  Repositories ({MSR}), {IEEE}, 2017, pp. 345--355.

\bibitem{DBLP:conf/msr/BellerBZJ14}
M.~Beller, A.~Bacchelli, A.~Zaidman, E.~J{\"{u}}rgens, Modern code reviews in
  open-source projects: which problems do they fix?, in: 11th Working
  Conference on Mining Software Repositories ({MSR}), {ACM}, 2014, pp.
  202--211.

\bibitem{DBLP:conf/icse/RigbyGS08}
P.~C. Rigby, D.~M. Germ{\'{a}}n, M.~D. Storey, Open source software peer review
  practices: a case study of the apache server, in: International Conference on
  Software Engineering {(ICSE}), {ACM}, 2008, pp. 541--550.

\bibitem{DBLP:conf/kbse/HiltonTHMD16}
M.~Hilton, T.~Tunnell, K.~Huang, D.~Marinov, D.~Dig, Usage, costs, and benefits
  of continuous integration in open-source projects, in: Proceedings of the
  31st {IEEE/ACM} International Conference on Automated Software Engineering
  ({ASE}), {ACM}, 2016, pp. 426--437.

\bibitem{DBLP:conf/wcre/CasseeVS20}
N.~Cassee, B.~Vasilescu, A.~Serebrenik, The silent helper: The impact of
  continuous integration on code reviews, in: Int'l Conference on Software
  Analysis, Evolution and Reengineering ({SANER}), {IEEE}, 2020, pp. 423--434.

\bibitem{DBLP:conf/kbse/ZhaoSZFV17}
Y.~Zhao, A.~Serebrenik, Y.~Zhou, V.~Filkov, B.~Vasilescu, The impact of
  continuous integration on other software development practices: a large-scale
  empirical study, in: Proceedings of the International Conference on Automated
  Software Engineering ({ASE}), {IEEE}, 2017, pp. 60--71.

\bibitem{DBLP:conf/wcre/ZampettiBCP19}
F.~Zampetti, G.~Bavota, G.~Canfora, M.~D. Penta, A study on the interplay
  between pull request review and continuous integration builds, in: 26th
  {IEEE} International Conference on Software Analysis, Evolution and
  Reengineering ({SANER}), {IEEE}, 2019, pp. 38--48.

\bibitem{DBLP:conf/wcre/PanichellaAPA15}
S.~Panichella, V.~Arnaoudova, M.~D. Penta, G.~Antoniol, Would static analysis
  tools help developers with code reviews?, in: 22nd {IEEE} International
  Conference on Software Analysis, Evolution, and Reengineering ({SANER}),
  {IEEE}, 2015, pp. 161--170.

\bibitem{DBLP:conf/icsm/NeryCK19}
G.~S. Nery, D.~A. da~Costa, U.~Kulesza, An empirical study of the relationship
  between continuous integration and test code evolution, in: 2019 {IEEE}
  International Conference on Software Maintenance and Evolution ({ICSME}),
  {IEEE}, 2019, pp. 426--436.

\bibitem{MantylaTSE2009}
M.~V. M{\"a}ntyl{\"a}, C.~Lassenius, What types of defects are really
  discovered in code reviews?, IEEE Transactions on Software Engineering 35~(3)
  (2009) 430--448.

\bibitem{BorgesJSS2018}
H.~Borges, M.~{Tulio Valente}, What's in a {GitHub} star? understanding
  repository starring practices in a social coding platform, Journal of Systems
  and Software 146 (2018) 112--129.

\bibitem{ali_khatami_2022_7404903}
A.~Khatami, A.~Zaidman,
  \href{https://doi.org/10.5281/zenodo.7404903}{{"State-Of-The-Practice in
  Quality Assurance in Java-Based Open Source Software Development" Replication
  Package}} (Dec. 2022).
\newblock \href {https://doi.org/10.5281/zenodo.7404903}
  {\path{doi:10.5281/zenodo.7404903}}.
\newline\urlprefix\url{https://doi.org/10.5281/zenodo.7404903}

\bibitem{DBLP:journals/smr/TufanoPBPOLP17}
M.~Tufano, F.~Palomba, G.~Bavota, M.~D. Penta, R.~Oliveto, A.~D. Lucia,
  D.~Poshyvanyk, There and back again: Can you compile that snapshot?, J.
  Softw. Evol. Process. 29~(4) (2017).

\bibitem{DBLP:journals/ese/Maes-BermejoGGR22}
M.~Maes{-}Bermejo, M.~Gallego, F.~Gort{\'{a}}zar, G.~Robles, J.~M.
  Gonz{\'{a}}lez{-}Barahona, Revisiting the building of past snapshots - a
  replication and reproduction study, Empir. Softw. Eng. 27~(3) (2022) 65.
\newblock \href {https://doi.org/10.1007/s10664-022-10117-6}
  {\path{doi:10.1007/s10664-022-10117-6}}.

\bibitem{DBLP:conf/esem/HassanMLW17}
F.~Hassan, S.~Mostafa, E.~S.~L. Lam, X.~Wang, Automatic building of java
  projects in software repositories: {A} study on feasibility and challenges,
  in: 2017 {ACM/IEEE} International Symposium on Empirical Software Engineering
  and Measurement ({ESEM}), {IEEE}, 2017, pp. 38--47.

\bibitem{DBLP:conf/esem/HassanW17}
F.~Hassan, X.~Wang, Change-aware build prediction model for stall avoidance in
  continuous integration, in: International Symposium on Empirical Software
  Engineering and Measurement ({ESEM}), {IEEE}, 2017, pp. 157--162.

\bibitem{DBLP:conf/msr/BellerGZ17a}
M.~Beller, G.~Gousios, A.~Zaidman, Travistorrent: synthesizing {Travis} {CI}
  and {GitHub} for full-stack research on continuous integration, in:
  Proceedings of the 14th International Conference on Mining Software
  Repositories ({MSR}), {IEEE}, 2017, pp. 447--450.

\bibitem{DBLP:conf/icse/JinS20}
X.~Jin, F.~Servant, A cost-efficient approach to building in continuous
  integration, in: Proc. International Conference on Software Engineering
  (ICSE), {ACM}, 2020, pp. 13--25.

\bibitem{DBLP:conf/sigsoft/LouCCHZ20}
Y.~Lou, Z.~Chen, Y.~Cao, D.~Hao, L.~Zhang, Understanding build issue resolution
  in practice: symptoms and fix patterns, in: Joint European Software
  Engineering Conference and Symposium on the Foundations of Software
  Engineering (ESEC/FSE), {ACM}, 2020, pp. 617--628.

\bibitem{DBLP:conf/msr/McIntoshKAH14}
S.~McIntosh, Y.~Kamei, B.~Adams, A.~E. Hassan, The impact of code review
  coverage and code review participation on software quality: a case study of
  the qt, vtk, and {ITK} projects, in: Proceedings of the 11th Working
  Conference on Mining Software Repositories ({MSR}), {ACM}, 2014, pp.
  192--201.

\bibitem{DBLP:conf/msr/RahmanR17}
M.~M. Rahman, C.~K. Roy, Impact of continuous integration on code reviews, in:
  Proc. International Conference on Mining Software Repositories ({MSR}),
  {IEEE}, 2017, pp. 499--502.

\bibitem{DBLP:journals/ibmsj/Fagen76}
M.~E. Fagan, Design and code inspections to reduce errors in program
  development, {IBM} Syst. J. 15~(3) (1976) 182--211.

\bibitem{DBLP:conf/icse/KononenkoBG16}
O.~Kononenko, O.~Baysal, M.~W. Godfrey, Code review quality: how developers see
  it, in: Proceedings of the International Conference on Software Engineering
  ({ICSE}), {ACM}, 2016, pp. 1028--1038.

\bibitem{DBLP:conf/icse/GopinathJG14}
R.~Gopinath, C.~Jensen, A.~Groce, Code coverage for suite evaluation by
  developers, in: 36th International Conference on Software Engineering
  ({ICSE}), {ACM}, 2014, pp. 72--82.
\newblock \href {https://doi.org/10.1145/2568225.2568278}
  {\path{doi:10.1145/2568225.2568278}}.

\bibitem{DBLP:conf/kbse/Hilton0M18}
M.~Hilton, J.~Bell, D.~Marinov, A large-scale study of test coverage evolution,
  in: Proceedings of the 33rd {ACM/IEEE} International Conference on Automated
  Software Engineering ({ASE}), {ACM}, 2018, pp. 53--63.

\bibitem{DBLP:conf/icsm/ElbaumGR01}
S.~G. Elbaum, D.~Gable, G.~Rothermel, The impact of software evolution on code
  coverage information, in: International Conference on Software Maintenance
  ({ICSM}), {IEEE}, 2001, pp. 170--179.

\bibitem{DBLP:journals/ese/ZaidmanRDD11}
A.~Zaidman, B.~{Van Rompaey}, A.~van Deursen, S.~Demeyer, Studying the
  co-evolution of production and test code in open source and industrial
  developer test processes through repository mining, Empir. Softw. Eng. 16~(3)
  (2011) 325--364.

\bibitem{DBLP:conf/icse/GousiosZSD15}
G.~Gousios, A.~Zaidman, M.~D. Storey, A.~van Deursen, Work practices and
  challenges in pull-based development: The integrator's perspective, in: 37th
  {IEEE/ACM} International Conference on Software Engineering (ICSE), {IEEE},
  2015, pp. 358--368.

\bibitem{DBLP:journals/tse/BellerGPPAZ19}
M.~Beller, G.~Gousios, A.~Panichella, S.~Proksch, S.~Amann, A.~Zaidman,
  Developer testing in the {IDE:} patterns, beliefs, and behavior, {IEEE}
  Trans. Software Eng. 45~(3) (2019) 261--284.

\bibitem{DBLP:journals/tr/KochharLLN17}
P.~S. Kochhar, D.~Lo, J.~Lawall, N.~Nagappan, Code coverage and postrelease
  defects: {A} large-scale study on open source projects, {IEEE} Trans. Reliab.
  66~(4) (2017) 1213--1228.
\newblock \href {https://doi.org/10.1109/TR.2017.2727062}
  {\path{doi:10.1109/TR.2017.2727062}}.

\bibitem{DBLP:journals/tse/AthanasiouNVZ14}
D.~Athanasiou, A.~Nugroho, J.~Visser, A.~Zaidman, Test code quality and its
  relation to issue handling performance, {IEEE} Trans. Software Eng. 40~(11)
  (2014) 1100--1125.

\bibitem{DBLP:conf/sigsoft/VasilescuYWDF15}
B.~Vasilescu, Y.~Yu, H.~Wang, P.~T. Devanbu, V.~Filkov, Quality and
  productivity outcomes relating to continuous integration in {GitHub}, in:
  Proceedings of the 2015 10th Joint Meeting on Foundations of Software
  Engineering ({ESEC/FSE}), {ACM}, 2015, pp. 805--816.

\bibitem{SonarSource}
Sonarqube, \url{https://www.sonarqube.org}, last visited May 20th, 2022.

\bibitem{DBLP:conf/kbse/VassalloPBG18}
C.~Vassallo, F.~Palomba, A.~Bacchelli, H.~C. Gall, Continuous code quality: are
  we (really) doing that?, in: Proceedings of the International Conference on
  Automated Software Engineering ({ASE}), {ACM}, 2018, pp. 790--795.

\bibitem{DBLP:conf/sigsoft/Hilton0TMD17}
M.~Hilton, N.~Nelson, T.~Tunnell, D.~Marinov, D.~Dig, Trade-offs in continuous
  integration: assurance, security, and flexibility, in: Proceedings of the
  2017 11th Joint Meeting on Foundations of Software Engineering ({ESEC/FSE}),
  {ACM}, 2017, pp. 197--207.
\newblock \href {https://doi.org/10.1145/3106237.3106270}
  {\path{doi:10.1145/3106237.3106270}}.

\bibitem{DBLP:conf/msr/GautamVS17}
A.~Gautam, S.~Vishwasrao, F.~Servant, An empirical study of activity,
  popularity, size, testing, and stability in continuous integration, in:
  Proceedings of the 14th International Conference on Mining Software
  Repositories ({MSR}), {IEEE} Computer Society, 2017, pp. 495--498.
\newblock \href {https://doi.org/10.1109/MSR.2017.38}
  {\path{doi:10.1109/MSR.2017.38}}.

\bibitem{DBLP:conf/icse/GousiosPD14}
G.~Gousios, M.~Pinzger, A.~van Deursen, An exploratory study of the pull-based
  software development model, in: 36th International Conference on Software
  Engineering ({ICSE}), {ACM}, 2014, pp. 345--355.

\bibitem{DBLP:conf/msr/GousiosZ14}
G.~Gousios, A.~Zaidman, A dataset for pull-based development research, in:
  Working Conf. on Mining Software Repositories ({MSR}), {ACM}, 2014, pp.
  368--371.

\bibitem{DBLP:conf/msr/ZhangR020}
X.~Zhang, A.~Rastogi, Y.~Yu, On the shoulders of giants: {A} new dataset for
  pull-based development research, in: {MSR} '20: 17th International Conference
  on Mining Software Repositories, {ACM}, 2020, pp. 543--547.

\bibitem{DBLP:conf/msr/KinsmanWGT21}
T.~Kinsman, M.~S. Wessel, M.~A. Gerosa, C.~Treude, How do software developers
  use {GitHub} actions to automate their workflows?, in: International
  Conference on Mining Software Repositories ({MSR}), {IEEE}, 2021, pp.
  420--431.

\bibitem{Dabic:msr2021data}
O.~Dabic, E.~Aghajani, G.~Bavota, Sampling projects in {GitHub} for {MSR}
  studies, in: International Conference on Mining Software Repositories
  ({MSR}), {IEEE}, 2021, pp. 560--564.

\bibitem{DBLP:journals/corr/abs-1712-01024}
M.~Sul{\'{\i}}r, J.~Porub{\"{a}}n, A quantitative study of java software
  buildability, CoRR abs/1712.01024 (2017).
\newblock \href {http://arxiv.org/abs/1712.01024} {\path{arXiv:1712.01024}}.

\bibitem{gallaba2022lessons}
K.~Gallaba, M.~Lamothe, S.~McIntosh, Lessons from eight years of operational
  data from a continuous integration service (2022).

\bibitem{wesselEMSE}
M.~Wessel, A.~Serebrenik, I.~Wiese, I.~Steinmacher, M.~A. Gerosa, Quality
  gatekeepers: Investigating the effects of code review bots on pull request
  activities, Empirical Software Engineering~\emph{To Appear}.

\bibitem{DBLP:conf/msr/KalliamvakouGBSGD14}
E.~Kalliamvakou, G.~Gousios, K.~Blincoe, L.~Singer, D.~M. Germ{\'{a}}n, D.~E.
  Damian, The promises and perils of mining github, in: P.~T. Devanbu, S.~Kim,
  M.~Pinzger (Eds.), 11th Working Conference on Mining Software Repositories
  ({MSR}), {ACM}, 2014, pp. 92--101.
\newblock \href {https://doi.org/10.1145/2597073.2597074}
  {\path{doi:10.1145/2597073.2597074}}.

\bibitem{DBLP:journals/sqj/HorvathGBTBG19}
F.~Horv{\'{a}}th, T.~Gergely, {\'{A}}.~Besz{\'{e}}des, D.~Tengeri, G.~Balogh,
  T.~Gyim{\'{o}}thy, Code coverage differences of java bytecode and source code
  instrumentation tools, Softw. Qual. J. 27~(1) (2019) 79--123.
\newblock \href {https://doi.org/10.1007/s11219-017-9389-z}
  {\path{doi:10.1007/s11219-017-9389-z}}.

\bibitem{DBLP:journals/ese/McIntoshNAMH15}
S.~McIntosh, M.~Nagappan, B.~Adams, A.~Mockus, A.~E. Hassan, A large-scale
  empirical study of the relationship between build technology and build
  maintenance, Empir. Softw. Eng. 20~(6) (2015) 1587--1633.

\bibitem{DBLP:conf/wcre/ThongtanunamMHI18}
P.~Thongtanunam, S.~McIntosh, A.~E. Hassan, H.~Iida, Review participation in
  modern code review: An empirical study of the android, qt, and openstack
  projects (journal-first abstract), in: 25th International Conference on
  Software Analysis, Evolution and Reengineering ({SANER}), {IEEE} Computer
  Society, 2018, p. 475.
\newblock \href {https://doi.org/10.1109/SANER.2018.8330241}
  {\path{doi:10.1109/SANER.2018.8330241}}.

\bibitem{DBLP:conf/icsm/AlvesYV10}
T.~L. Alves, C.~Ypma, J.~Visser, Deriving metric thresholds from benchmark
  data, in: 26th {IEEE} International Conference on Software Maintenance
  {(ICSM}), {IEEE}, 2010, pp. 1--10.

\bibitem{Zhu1997}
H.~Zhu, P.~A.~V. Hall, J.~H. May, Software unit test coverage and adequacy, ACM
  Computing Surveys 29~(4) (1997).

\bibitem{DBLP:conf/quatic/HeitlagerKV07}
I.~Heitlager, T.~Kuipers, J.~Visser, A practical model for measuring
  maintainability, in: Quality of Information and Communications Technology,
  6th International Conference on the Quality of Information and Communications
  Technology ({QUATIC}), {IEEE} Computer Society, 2007, pp. 30--39.
\newblock \href {https://doi.org/10.1109/QUATIC.2007.8}
  {\path{doi:10.1109/QUATIC.2007.8}}.

\bibitem{dekking2005modern}
F.~M. Dekking, C.~Kraaikamp, H.~P. Lopuha{\"a}, L.~E. Meester, A Modern
  Introduction to Probability and Statistics: Understanding why and how, Vol.
  488, Springer, 2005.

\bibitem{DBLP:journals/data/SulirBMCJ20}
M.~Sul{\'{\i}}r, M.~Bac{\'{\i}}kov{\'{a}}, M.~Madeja, S.~Chodarev,
  J.~Juh{\'{a}}r, Large-scale dataset of local java software build results,
  Data 5~(3) (2020) 86.
\newblock \href {https://doi.org/10.3390/data5030086}
  {\path{doi:10.3390/data5030086}}.

\bibitem{DBLP:conf/kbse/Hassan19}
F.~Hassan, Tackling build failures in continuous integration, in: 34th
  {IEEE/ACM} International Conference on Automated Software Engineering
  ({ASE}), {IEEE}, 2019, pp. 1242--1245.
\newblock \href {https://doi.org/10.1109/ASE.2019.00150}
  {\path{doi:10.1109/ASE.2019.00150}}.

\bibitem{DBLP:conf/wcre/MachoM018}
C.~Macho, S.~McIntosh, M.~Pinzger, Automatically repairing dependency-related
  build breakage, in: 25th International Conference on Software Analysis,
  Evolution and Reengineering ({SANER}), {IEEE} Computer Society, 2018, pp.
  106--117.
\newblock \href {https://doi.org/10.1109/SANER.2018.8330201}
  {\path{doi:10.1109/SANER.2018.8330201}}.

\bibitem{DBLP:journals/jss/BarrakEAK21}
A.~Barrak, E.~E. Eghan, B.~Adams, F.~Khomh, Why do builds fail? - {A}
  conceptual replication study, J. Syst. Softw. 177 (2021) 110939.
\newblock \href {https://doi.org/10.1016/j.jss.2021.110939}
  {\path{doi:10.1016/j.jss.2021.110939}}.

\bibitem{DBLP:conf/icsm/KononenkoBGCG15}
O.~Kononenko, O.~Baysal, L.~Guerrouj, Y.~Cao, M.~W. Godfrey, Investigating code
  review quality: Do people and participation matter?, in: 2015 {IEEE}
  International Conference on Software Maintenance and Evolution ({ICSME}),
  {IEEE} Computer Society, 2015, pp. 111--120.
\newblock \href {https://doi.org/10.1109/ICSM.2015.7332457}
  {\path{doi:10.1109/ICSM.2015.7332457}}.

\bibitem{DBLP:conf/wcre/VassalloPPPZG18}
C.~Vassallo, S.~Panichella, F.~Palomba, S.~Proksch, A.~Zaidman, H.~C. Gall,
  Context is king: The developer perspective on the usage of static analysis
  tools, in: 25th International Conference on Software Analysis, Evolution and
  Reengineering, ({SANER}), {IEEE} Computer Society, 2018, pp. 38--49.
\newblock \href {https://doi.org/10.1109/SANER.2018.8330195}
  {\path{doi:10.1109/SANER.2018.8330195}}.

\bibitem{DBLP:conf/msr/ZampettiSOCP17}
F.~Zampetti, S.~Scalabrino, R.~Oliveto, G.~Canfora, M.~D. Penta, How open
  source projects use static code analysis tools in continuous integration
  pipelines, in: Proceedings of the 14th International Conference on Mining
  Software Repositories ({MSR}), {IEEE} Computer Society, 2017, pp. 334--344.
\newblock \href {https://doi.org/10.1109/MSR.2017.2}
  {\path{doi:10.1109/MSR.2017.2}}.

\bibitem{golzadehrise}
M.~Golzadeh, A.~Decan, T.~Mens, On the rise and fall of {CI} services in
  {GitHub}, in: 29th IEEE International Conference on Software Analysis,
  Evolution and Reengineering ({SANER}), {IEEE}, 2022.

\bibitem{DBLP:conf/msr/BernardoCK18}
J.~H. Bernardo, D.~A. da~Costa, U.~Kulesza, Studying the impact of adopting
  continuous integration on the delivery time of pull requests, in: Proceedings
  of the 15th International Conference on Mining Software Repositories ({MSR}),
  {ACM}, 2018, pp. 131--141.

\bibitem{DBLP:conf/sigsoft/RahmanAKS18}
A.~Rahman, A.~Agrawal, R.~Krishna, A.~Sobran, Characterizing the influence of
  continuous integration: empirical results from 250+ open source and
  proprietary projects, in: Proceedings of the 4th {ACM} {SIGSOFT}
  International Workshop on Software Analytics, SWAN@ESEC/SIGSOFT {FSE}, {ACM},
  2018, pp. 8--14.
\newblock \href {https://doi.org/10.1145/3278142.3278149}
  {\path{doi:10.1145/3278142.3278149}}.

\bibitem{DBLP:conf/msr/BosuGB15}
A.~Bosu, M.~Greiler, C.~Bird, Characteristics of useful code reviews: An
  empirical study at microsoft, in: 12th {IEEE/ACM} Working Conference on
  Mining Software Repositories ({MSR}), {IEEE}, 2015, pp. 146--156.

\bibitem{DBLP:journals/tse/DoWA22}
L.~N.~Q. Do, J.~R. Wright, K.~Ali, Why do software developers use static
  analysis tools? {A} user-centered study of developer needs and motivations,
  {IEEE} Trans. Software Eng. 48~(3) (2022) 835--847.
\newblock \href {https://doi.org/10.1109/TSE.2020.3004525}
  {\path{doi:10.1109/TSE.2020.3004525}}.

\bibitem{DBLP:journals/jss/WangMLM22}
Y.~Wang, M.~V. M{\"{a}}ntyl{\"{a}}, Z.~Liu, J.~Markkula,
  \href{https://doi.org/10.1016/j.jss.2022.111259}{Test automation maturity
  improves product quality - quantitative study of open source projects using
  continuous integration}, J. Syst. Softw. 188 (2022) 111259.
\newblock \href {https://doi.org/10.1016/j.jss.2022.111259}
  {\path{doi:10.1016/j.jss.2022.111259}}.
\newline\urlprefix\url{https://doi.org/10.1016/j.jss.2022.111259}

\end{thebibliography}

\end{document}